\documentclass[prd,nofootinbib,floats,superscriptaddress,eqsecnum,tightenlines,preprintnumbers,11pt]{revtex4}

\usepackage{hyperref}
\usepackage{graphicx}
\usepackage{amsmath,amssymb,amsfonts,amsthm,latexsym}
\usepackage{marginnote}
\usepackage{color}

\def\be{\begin{equation}}
\def\ee{\end{equation}}
\def\beq{\begin{equation}}
\def\eeq{\end{equation}}
\def\bi{\begin{itemize}}
\def\ei{\end{itemize}}
\def\ba{\begin{array}}
\def\ea{\end{array}}
\def\bfig{\begin{figure}}
\def\efig{\end{figure}}

\newcommand{\bea}{\begin{eqnarray}}
\newcommand{\eea}{\end{eqnarray}}
\newcommand\dd{\mathrm d}

\newcommand{\mn}{\mathfrak{n}}
\newcommand{\ma}{\mathfrak{a}}

\begin{document}

\title{Generalisation of Conformal-Disformal Transformations \\
of \\
the Metric in Scalar-Tensor  Theories}

\author{Eugeny Babichev}
\affiliation{Laboratoire de Physique des deux Infinis IJCLab, Universit\'e Paris-Saclay, CNRS, France}
\author{Keisuke Izumi}
\affiliation{Department of Mathematics, Nagoya University, Nagoya 464-8602, Japan}
\affiliation{Kobayashi-Maskawa Institute, Nagoya University, Nagoya 464-8602, Japan}

\author{Karim Noui}
\affiliation{Laboratoire de Physique des deux Infinis IJCLab, Universit\'e Paris-Saclay, CNRS, France}

\author{Norihiro Tanahashi}
\affiliation{Department of Physics, Kyoto University, Kyoto 606-8502, Japan}

\author{Masahide Yamaguchi}
\affiliation{Cosmology, Gravity, and Astroparticle Physics Group, Center for Theoretical Physics of the Universe, Institute for Basic Science (IBS), Daejeon, 34126, Korea}
\affiliation{Department of Physics, Tokyo Institute of Technology, Tokyo 152-8551, Japan}

\date{\today}

\preprint{KUNS-3003}

\begin{abstract}
We study new classes of metric transformations  in the context of scalar-tensor theories, which involve both higher derivatives of the scalar field and  derivatives of the metric itself. In general, such transformations are not invertible as they involve derivatives of the metric,
which typically leads to instability due to Ostrogradsky ghosts.
We show, however, that a certain class of this type of transformations is invertible: we construct new examples of invertible conformal (and also disformal) transformations with higher derivatives. 
Finally, we make use of these new transformations to construct extended mimetic theories of gravity, and we study 
their properties in the context of cosmology.
\end{abstract}

\maketitle

\tableofcontents

\section{Introduction}
The discovery of the accelerated expansion of the universe has catalysed the construction and study of alternative theories of gravitation. While the construction of these theories was initially motivated by the idea of finding alternatives to the cosmological constant, it very quickly opened up a new field of research into modified theories of gravitation, with applications in cosmology and astrophysics (see \cite{Berti:2015itd} and references therein). Now that we know how to detect gravitational waves~\cite{LIGOScientific:2016aoc}, modified  theories of gravity are being used in particular to  test  general relativity  in the strong gravity regime. 

In this article,
we mainly examine formal properties of modified theories of gravity.
We will consider 
scalar-tensor theories in which a scalar field 
is a part of gravitational interaction along with the metric.
They are therefore seen as effective theories whose differences with general relativity could come from new physics in the ultraviolet or eventually in the infrared. Scalar-tensor theories have been studied extensively, especially over the last twenty years, when it was realised that introducing higher-order derivatives (of the scalar field and the metric) into the Lagrangian led to very rich structures and an 
interesting phenomenology~\cite{Dvali:2000hr,Nicolis:2008in}. At the same time, introducing higher-order derivatives 
may give rise to the well-known Ostrogradsky instabilities that would render the theory obsolete \cite{Ostrogradsky:1850fid,Woodard:2015zca}. This led to an effort to classify tensor-scalar theories that do not exhibit such instabilities, resulting in the classification of DHOST theories \cite{Langlois:2015cwa,Langlois:2015skt,Langlois:2018jdg,Achour:2016rkg,BenAchour:2016fzp,Crisostomi:2016czh}. In this classification, the notion of disformal transformation  of the metric~\cite{Zumalac_rregui_2014} plays a crucial role, as they allow us to construct equivalent classes of DHOST theories, 
and to introduce non-minimal (derivative) couplings to matter. 
Furthermore,  disformal transformations are an effective tool for constructing 
new solutions, such as black holes or other exotic objects 
(see \cite{Achour:2021pla,Anson:2020trg,BenAchour:2020fgy,BenAchour:2020wiw} for instance). 

In this paper, we introduce a new class of metric transformations which generalises disformal transformations.  The question of generalising disformal transformations has already been studied recently, leading to 
a study of transformations of the metric that involve not only derivatives of the scalar field but also of the metric itself \cite{Takahashi_2022,Takahashi:2023jro,Takahashi_2023}.
In general, such transformations are not invertible and typically generate Ostrogradsky ghosts, which makes them of limited interest. Indeed, one needs to ``integrate'' a partial differential equation to ``invert'' the relation and to express the original metric in terms of the transformed metric. Such an integration introduces necessarily integration constants, rendering a transformation to be non-invertible.
However, 
a subclass of such transformation is invertible, as it has been shown in ~\cite{Babichev:2019twf, Babichev:2021bim, Takahashi:2023vva,Takahashi_2022,Takahashi_2023}, where interesting examples of higher derivative transformations of the metric have been exhibited.
Here, we generalise their construction to build a new class of higher order conformal transformations of the metric. 

In order to demonstrate the method 
we start with a simple toy model defined by three free particles, which mimic 2 polarisations of a  graviton and 1 scalar-like particle, in the context of classical mechanics. Then, we introduce two classes of transformations of the dynamical variables that we call algebraic transformations and differential transformations. The former are standard: the new variables are (non-linear) functions
of the original ones and the transformation is invertible when the Jacobian of the transformation is itself 
non-degenerate.
The latter case is more interesting: the new variables are (non-linear) functions
of the original ones and their derivatives. In general, these transformations are not invertible, however one can exhibit an invertible  subclass. We find  sufficient conditions\footnote{Necessary conditions have also been discussed in \cite{Babichev:2019twf,Babichev:2021bim} in a slightly different context.} for the transformations to be invertible
and construct explicit examples.
We then study the outcome of applying both invertible and non-invertible transformations for the dynamics of the particles. 
In particular, we discuss whether 
they introduce new degrees of freedom and whether these extra degrees of freedom are Ostrogradsky ghosts. 

Finally,  we adapt and extend the method 
to scalar-tensor theories and we introduce a new class of generalised
conformal transformations of the metric using the concept of the mimetic metric. We discuss the possibility to extend this new class to higher derivative disformal transformations. As an application, we make use of these new conformal transformations to define extended mimetic theories and to study 
their properties.

The paper is organised as follows. In section~\ref{sec:toymodel}, we introduce a toy model which describes the dynamics of three particles, and we consider transformations of the dynamical variables which mimic those of the metric in scalar-tensor theories. We study the conditions for these transformations to be invertible and the consequences (in terms of degrees of freedom) of these transformations for the dynamics of the particles. We illustrate the results with concrete examples. In section~\ref{sec:gentrsf-metric}, we consider higher derivative transformations of the metric and introduce a new class of such transformations which are shown to be invertible. We first focus on higher derivative conformal transformations before generalising to disformal transformations. The construction of these transformations deeply relies on the notion of mimetic metric that we 
review in this section. Then, we make use of these new transformations to define extended mimetic theories of gravity. We examine the properties of the resulting new theories, by restricting to the case of a flat Friedmann-Lema\^itre-Robertson-Walker (FLRW) metric. 
Some of the technical details are provided in the appendices. 
Appendix~\ref{AppA} discusses some generalisations of the toy models and their relevance to the standard/extended mimetic theories. Appendix~\ref{App:derive_C1} presents an alternative derivation of the invertible conformal transformations with higher derivatives discussed in section~\ref{sec:gentrsf-metric}. Appendix~\ref{AppB} shows the detail of the Hamiltonian analysis for the extended mimetic theory on the FLRW spacetime, and contrasts it with the result for the standard mimetic theory.

\section{Toy Model from classical mechanics}
\label{sec:toymodel}

In this section, we consider a simple toy model which mimics, in the context of classical mechanics, the dynamics of scalar-tensor theories. Such a simple model will enable us to 
introduce and to illustrate some properties of metric transformations avoiding complex technical aspects.

\subsection{Definition of the model}

We introduce two variables $g_1(t)$ and $g_2(t)$ that mimic two graviton polarisations along with a scalar-like variable $\phi(t)$. To start with, we assume that they are free particles and then, their dynamics are governed by the free action
\begin{eqnarray}
\label{freeaction}
S_0[g_1,g_2,\phi] \; = \; \frac{1}{2} \int \, \dd t \, \left( \dot{g_1}^2 + \dot{g_2}^2  + \dot\phi^2\right) \, .
\label{S_0}
\end{eqnarray}
One can add a potential to make the dynamics less trivial but this will not be relevant for our discussion here. 
The space of solutions, which is parametrised by $(g_1,g_2,\phi)$ at any time, is usually  called the configuration space of the system.

Next, we consider a different dynamics for the three variables $g_1$, $g_2$  and $\phi$ which is governed by the  action
\begin{eqnarray}
\label{transfoactionmodel}
S[g_1,g_2,\phi] \; =\; S_0[\tilde g_1, \tilde g_2,\phi] \,,
\end{eqnarray}
where the ``new variables'' $\tilde g_1$ and $\tilde g_2$ are linked to $g_1, g_2$ and $\phi$ by  general relations of the form,
\begin{eqnarray}
\label{generaltransfo}
\tilde g_n \; = \; F_n (g_1, g_2 , \phi \, ; \, \dot g_1 , \dot g_2 \, , \dot \phi) \, .
\end{eqnarray}
Hence the two functions $F_n$ could depend not only on the  variables $g_i(t)$ and $\phi(t)$ but also 
on their derivatives. In this new theory, $g_n$ and $\phi$ are not in general free particles anymore. 

However, when the transformation \eqref{generaltransfo} is invertible in the sense that  one can uniquely express $\tilde g_1$ and $\tilde g_2$ in terms of the variables $g_i$, $\phi$ and their derivatives (without need of an integration), the theory is equivalent to the original free particle theory and propagates only three degrees of freedom. 

At this stage, the property of the transformation to be invertible is rather abstract and it needs to be refined.
To do this, it is convenient to separate the transformations into two classes that we call algebraic and differential transformations. 
Then we illustrate the result of a non-invertible  transformation  providing  a simple example. 

\subsection{Algebraic Transformations}
First, we consider algebraic transformations which are defined by the fact that the functions $F_1$ and $F_2$ depend on the variables $g_1$ and $g_2$ only, and not on their derivatives. Notice that they can also depend on $\phi$ and $\dot\phi$, so that
\begin{eqnarray}
\label{generaltransfoalgebraic} 
\tilde g_n \; = \; F_n (g_1, g_2 , \phi \, ; \dot \phi) \, .
\end{eqnarray}
 When the transformation \eqref{generaltransfoalgebraic}  is not linear in the variables $g_n$, the question on its invertibility might be addressed, in general, only locally, i.e.\ in an open set at the vicinity of a point $(g_1,g_2)$ of the configuration space.\footnote{As the dependency on $\phi$ and $\dot \phi$ is not relevant for defining the invertibility of the transformation,
 we will omit to mention these variables when it is not needed for simplicity.}
According to the implicit function theorem, the transformation \eqref{generaltransfoalgebraic} is invertible in an open set at the vicinity of the point $(g_1,g_2)$  when the Jacobian matrix $J(g_1,g_2)$ is (locally) invertible with,
\begin{eqnarray}
\label{Jacob}
J(g_1,g_2) \; = \;
\left(\begin{array}{cc} 
\partial F_1/\partial g_1 & \partial F_1/\partial g_2 \\ 
\partial F_2/\partial g_1 & \partial F_2/\partial g_2
\end{array} 
\right) \, .
\end{eqnarray} 
In that case, the theory defined by the action \eqref{transfoactionmodel} obviously still propagates three degrees of freedom.

We will show that in the case of non-invertible transformations,  the number and the nature of the degrees of freedom in the theory depend on whether $F_1$ and
$F_2$ are functions of $\dot \phi$ or not.  To see more precisely how this works, we express the action \eqref{transfoactionmodel} in terms of $g_1$, $g_2$, $\phi$ and their (up to second) derivatives, and we make a Hamiltonian analysis. 

\subsubsection{Equivalent action and phase space}
To absorb second derivatives of the scalar field in the action, we proceed in a standard manner by introducing a new variable $A_*$ and a Lagrange multiplier. The original action is equivalent to the action below, which does not involve explicit higher derivatives,
\begin{eqnarray}
\label{eqaction}
S_{\textrm{eq}} [g_1,g_2,\phi,A_*] \; = \;  \int \dd t \, \left[ \frac{1}{2} \left( V_1^2 + V_2^2 + A_*^2 \right) +  \pi_\phi(\dot \phi - A_*) \right] \,.
\end{eqnarray}
Here we introduced the Lagrange multiplier $\pi_\phi$ enforcing the condition $A_*=\dot \phi$, and the quantities $V_n$ which are defined by the relation,
\begin{eqnarray}
\begin{pmatrix}
V_1 \\ V_2
\end{pmatrix}
=
\begin{pmatrix}
\dot{\tilde g}_1 \\ \dot{\tilde g}_2
\end{pmatrix}
=
J 
\begin{pmatrix}
\dot g_1 \\ \dot g_2
\end{pmatrix}
+
\begin{pmatrix}
J_{1*} \\ J_{2*}
\end{pmatrix} \dot A_*
+
\begin{pmatrix}
J_{1\phi} \\ J_{2\phi}
\end{pmatrix} A_* \, ,
\end{eqnarray}
where $J$ is the Jacobian matrix {\eqref{Jacob}} while
\begin{eqnarray}
J_{n *} = \frac{\partial F_n}{\partial \dot \phi} \, , \qquad
J_{n \phi} = \frac{\partial F_n}{\partial \phi} \, ,
\end{eqnarray}
which are understood as functions of $(g_1, g_2, A_*, \phi)$ (but not $\dot \phi$).
We are now ready to make a Hamiltonian analysis of the theory. We start by introducing the momenta associated with the variables $g_n$, $A_*$ and $\phi$,
\begin{eqnarray}
\label{phasespace}
\{ g_n \, , \, \pi_n\} = 1 \, , \qquad
\{ A_* , \pi_*\} = 1 \, , \qquad
\{ \phi , \pi_\phi \} = 1 \, , 
\end{eqnarray}
as it is clear that $\pi_\phi$ is the momenta associated to $\phi$.
In order to compute the Hamiltonian, we express the two remaining momenta in terms of the velocities,
\begin{eqnarray}
\label{momenta}
\begin{pmatrix} 
\pi_1 \\
\pi_2
\end{pmatrix} = {}^t \!J \, \begin{pmatrix} V_1 \\ V_2 \end{pmatrix} \, , \qquad
\pi_* = J_{1*} V_1 + J_{2*} V_2 \, ,
\end{eqnarray}
where ${}^t \!J$ is the transpose of $J$.
To go further, we need to treat separately the cases where $J$ is invertible and where it is not. 

\subsubsection{Invertible algebraic transformations}
In the former, as we can express $V_1$ and $V_2$ in terms of $\pi_1$ and $\pi_2$, the last relation in \eqref{momenta} leads to a primary constraint  which can be written in a matrix form as follows
\begin{eqnarray}
\chi \; = \; \pi_* - \begin{pmatrix} \pi_1 & \pi_2 \end{pmatrix} J^{-1}  \begin{pmatrix} J_{1*}  \\ J_{2*}  \end{pmatrix} \, \simeq \, 0 \, .
\end{eqnarray}
Hence, the total Hamiltonian takes the form
\begin{eqnarray}
\label{totalH}
H \; = \; H_0 + \mu \, \chi \, , 
\end{eqnarray}
where the canonical Halmiltonian $H_0$ is given by,
\begin{eqnarray} 
\label{canonicalH0}
H_0 = \frac{1}{2} \begin{pmatrix} \pi_1 & \pi_2 \end{pmatrix} ({}^t \! J J)^{-1}  \begin{pmatrix} \pi_1 \\ \pi_2 \end{pmatrix} -  \begin{pmatrix} \pi_1 & \pi_2 \end{pmatrix} J^{-1}  \begin{pmatrix} J_{1\phi} \\ J_{2\phi} \end{pmatrix} A_* - \frac{1}{2} A_*^2 + A_* \pi_\phi \, ,
\end{eqnarray}
and $\mu$ is a Lagrange multiplier which enforces the primary constraint. Requiring the stability of the constraint $\chi$ leads generically\footnote{\label{foot}Under specific conditions on the functions $F_1$ and $F_2$, $\chi$ could be first class. In that case, its time evolution could lead to a new first class constraint or to no constraint at all. We do not consider such particular situations here.} to a secondary constraint of the form,
\begin{eqnarray}
\psi = \{ \chi,H \} \simeq  \{  \chi,H_0  \}  \simeq -\pi_\phi \, + \, {\cal P} \simeq 0 \, ,
\end{eqnarray}
where $\cal P$ is a function of the phase space variables $(\pi_n,g_n,A_*,\phi) $ which can be computed explicitly, but its form is not needed here.  
 In general (see footnote \ref{foot} below), the Dirac analysis stops with a fixing of the Lagrange multiplier $\mu$ from the requirement that $\psi$ is stable under time evolution, i.e.\ $\{ H , \psi \} \simeq 0$. In that generic situation, the two constraints are second class, thus they enable us to eliminate two degrees of freedom in the phase space so that at the end the theory propagates 3 degrees of freedom, as expected.  Furthermore,  the constraint $\psi \simeq 0$ can be used to get rid of  $\pi_\phi$ and then the Hamiltonian \eqref{canonicalH0} is no more linear in $\pi_\phi$ which would have been responsible of the infamous Ostrogradsky ghost. As a consequence, none of the three degrees of freedom corresponds to an Ostrogradsky ghost. This is consistent with our expectation when we  consider  an invertible algebraic transformation.  

\subsubsection{Non-invertible algebraic transformations}

Let us now study the case where the transformation is  not invertible which means that the Jacobian matrix {\eqref{Jacob}} itself is not invertible.\footnote{Notice that examples have been found where, interestingly, the transformation is still invertible even though the Jacobian matrix is not invertible, and it can change the number of dynamical degrees of freedom in \cite{Jirousek:2022jhh}. However, in this paper, we assume that the determinant of the Jacobian matrix vanishes as a function while, in the particular example, the Jacobian matrix is not invertible only at some points.} The Hamiltonian analysis is exactly the same as in the previous case up to the computation of the momenta \eqref{momenta}. Indeed, if $J$ is not invertible but non-vanishing, it admits 
one null vector whose components are denoted 
$(n_1 \; n_2)$, i.e.\ 
\begin{eqnarray}
(n_1 \, n_2) \cdot {}^t \! J = 0 \, . 
\end{eqnarray} 
Let us see the consequences of the non-invertibility of $J$. First of all, we can always choose the components  $(n_1 \; n_2)$ so that  the matrix $J$ is of the form
\begin{eqnarray}
\label{degnerateJacob}
J = \begin{pmatrix} n_2 & - n_1 \\  \lambda n_2 & - \lambda n_1 \end{pmatrix} \, \quad \text{or} \quad
J = \begin{pmatrix} \lambda n_2 & - \lambda n_1 \\  n_2 & - n_1 \end{pmatrix} \, ,
\end{eqnarray}
where $\lambda$ is a function of the variables $(g_1,g_2,A_*,\phi)$ which could be null. Such Jacobian matrices correspond to  transformations \eqref{generaltransfoalgebraic} where either 
\begin{eqnarray}
\label{condnoninvert}
F_2= G(F_1,\phi,\dot\phi)  \, \quad \text{or} \quad
F_1=G(F_2,\phi,\dot\phi) \, ,
\end{eqnarray}
with $G$ being an arbitrary function, if $J$ takes the form of the l.h.s.\ or the r.h.s.\ of \eqref{degnerateJacob}.
It is easy to see that $\lambda = \partial G/\partial F_1$ in the former or $\lambda = \partial G/\partial F_2$ in the latter. 

Next we make the canonical analysis. We start from the phase space variables~\eqref{phasespace} together with the expressions of the momenta~\eqref{momenta}. In that case, the three equations \eqref{momenta} lead to a different primary constraint, compared to the previous situation, which is now of the form
\begin{eqnarray}
\chi \; = \; n_1 \pi_1 + n_2 \pi_2 \; \simeq \; 0 \, .
\end{eqnarray}
It can be used to eliminate the momentum $\pi_2$ for instance. Hence, the last equation of~\eqref{momenta} is no longer a primary constraint and this changes 
drastically the conclusions of the previous analysis. 
Indeed, one can immediately compute  the total Hamiltonian which is of the form \eqref{totalH} with a total Hamiltonian linear in the variable $\pi_\phi$ according to,
\begin{eqnarray}
H_0 \; = \; Q(\pi_1,\pi_*) + A_* \pi_\phi \, ,
\end{eqnarray}
where $Q$ is a polynomial at most quadratic in the momenta $\pi_1$ and $\pi_*$ whose expression is not needed. Notice that, to simplify notations, we omitted to mention the dependency of $Q$ in the other phase space variables. 
Contrary to what happens in the case of an invertible transformation, the stability of $\chi$ under time evolution does not enable us to solve $\pi_\phi$ anymore. As a consequence, generically, the theory still propagates 3 degrees of freedom but the Hamiltonian is not bounded from below nor from above. Hence, one of the degrees of freedom is an Ostrogradsky ghost. 

\medskip

In order to be more explicit and precise, we perform again the canonical analysis but differently, starting from the action \eqref{transfoactionmodel} in the case where $F_2=G(F_1,\phi,\dot \phi)$ (the other case~\eqref{condnoninvert} can be treated the same way). Hence, the variable $g_2$ can be eliminated and the action depends now
on the two independent variables $g_1$ and $\phi$. Furthermore, as $g_1$ enters the action through the function $F_1$ only, it is convenient to change variable and to 
consider 
$\mathcal{F}=F_1$ 
instead of $g_1$. Finally, the dynamics of the system is  described by the action
\begin{eqnarray}
S[\mathcal{F},\phi] = \frac{1}{2} \int \dd{t} \left[ \dot {\mathcal{F}}^2 + \left( \frac{\partial G}{\partial \mathcal{F}} \,\dot {\mathcal{F}} + \frac{\partial G}{\partial \phi} \,\dot\phi + \frac{\partial G}{\partial \dot\phi}\,\ddot \phi \right)^2 + \dot\phi^2\right] \, .
\end{eqnarray}
Now, the analysis of this action is straightforward. If the function $G$ involves $\dot \phi$, then the theory propagates three degrees of freedom, one of them being an Ostrogradsky ghost. However, when $G$ does not depend on $\dot \phi$ and, even if the functions $F_n$ depend on $\dot \phi$, the theory propagates two degrees of freedom with no Ostrogradsky ghost.

\subsection{Differential Transformations}
Now, we consider differential transformations where the functions $F_1$ and $F_2$ depend on the variables $g_1$ and $g_2$ and also on their derivatives $\dot g_1$ and $\dot g_2$. They could also depend on $\phi$ and $\dot\phi$. 

\subsubsection{Notations and general results}
In general, if one wants to ``invert'' the transformation \eqref{generaltransfo},
\begin{eqnarray}
\label{derivativetransfo}
\tilde g_n \; = \; F_n (g_1, g_2 \, ; \, \dot g_1 , \dot g_2 \, ; \, \phi,\dot \phi) \, , 
\end{eqnarray}
i.e.\ to express $g_n$ in terms of $\tilde g_n$, one has to  integrate two differential equations. Such an integration requires extra integration constants,  meaning extra degrees
of freedom in the theory.  Therefore, if expressing $g_n$ in terms of $\tilde g_n$ requires integrations, 
 the transformation is not invertible and the theory whose
dynamics is governed by the action 
\begin{eqnarray}
S[g_1,g_2,\phi] := S_0[\tilde g_1 (g_1, g_2 \, ; \, \dot g_1 , \dot g_2 \, ; \, \phi,\dot \phi), \tilde g_2 (g_1, g_2 \, ; \, \dot g_1 , \dot g_2 \, ; \, \phi,\dot \phi),\phi]
\end{eqnarray}
propagates extra degrees of freedom in comparison to $S_0[g_1,g_2,\phi]$. These extra degrees of freedom are expected to be Ostrogradsky ghosts because the action 
involves higher derivatives of the variables, as shown in the following expression
\begin{eqnarray}
S[g_1,g_2,\phi] \; = \;   \int \dd{t} \; L[\ddot g_1,\dot g_1,g_1; \ddot g_2, \dot g_2,g_2 ; \ddot \phi, \dot\phi,\phi] \, ,
\end{eqnarray}
where the Lagrangian given by,
\begin{eqnarray}
L= \frac{1}{2} \left[ \sum_{n=1}^2 \left( \frac{\partial F_n}{\partial \dot g_1} \ddot g_1 + \frac{\partial F_n}{\partial \dot g_2} \ddot g_2 + \frac{\partial F_n}{\partial g_1} \dot g_1 + \frac{\partial F_n}{\partial g_2} \dot g_2+  \frac{\partial F_n}{\partial \dot\phi} \ddot \phi + \frac{\partial F_n}{\partial \phi} \dot \phi \right)^2 + \dot\phi^2 \right] \, .
\end{eqnarray}
To study this theory and its number of degrees of freedom, we proceed as usual and introduce the equivalent action 
\begin{eqnarray}
S_{\rm{eq}}\; = \;   \int \dd{t} \left[ \frac{1}{2} 
\left( V_1^2 + V_2^2 + 
A_*^2
\right)
+ \pi_1(\dot g_1 - A_{*1}) + \pi_2(\dot g_2 - A_{*2}) + \pi_\phi(\dot \phi - A_{*}) \right]  \, ,
\end{eqnarray}
where $V_n$ have been defined by,
\begin{eqnarray} 
V_n 
= \dot {\tilde g}_n
=  \frac{\partial F_n}{\partial A_{*1}} \dot A_{*1} + \frac{\partial F_n}{\partial A_{*2}} \dot A_{*2} + \frac{\partial F_n}{\partial g_1} A_{*1} + \frac{\partial F_n}{\partial g_2} A_{*2}+  \frac{\partial F_n}{\partial A_*} \dot A_* + \frac{\partial F_n}{\partial \phi} A_*  \, .\label{defofVn}
\end{eqnarray}
Higher derivatives of the variables have been absorbed into the new variables $A_{*n}=\dot g_n$ and $A_*=\dot \phi$ by virtue of Lagrange multipliers. 

At this stage the phase space is parametrised by
the following 6 pairs of variables,
\begin{eqnarray}
\label{pairofcanon}
\{ g_n \, , \, \pi_n\} = 1 \, , \qquad
\{ A_{*n} \, , \, \pi_{*n}\} = 1 \, , \qquad
\{ A_* , \pi_*\} = 1 \, , \qquad
\{ \phi , \pi_\phi \} = 1 \, .
\end{eqnarray}
The calculation of the momenta $\pi_{*n}$ and $\pi_*$ is immediate and leads to 
\begin{eqnarray}
\label{momentadifftransfo}
\pi_{*n} = \frac{\partial F_1}{\partial A_{*n}} V_1 + \frac{\partial F_2}{\partial A_{*n}} V_2 \, , \qquad \pi_* = \frac{\partial F_1}{\partial A_*} V_1 + \frac{\partial F_2}{\partial A_*} V_2 \, .
\end{eqnarray}
Therefore, these three momenta are not independent and the theory admits necessarily at least one primary constraint  of the form
\begin{eqnarray}
\label{primarychi}
\alpha_1 \, \pi_{*1} \, + \,   \alpha_2 \, \pi_{*2} \, + \, \alpha_\phi \, \pi_* \, \simeq \, 0 \;  ,
\end{eqnarray}
where $\alpha_{n*}$ and $\alpha_\phi$ are functions of the phase space variables that can be constructed from the functions $F_n$ and their derivatives. 
The theory could have up to 3 primary constraints of that type.

If there are 3 primary constraints, this means that $\pi_{*n}$ and $\pi_*$ are vanishing, therefore the theory does not involve higher derivatives, hence  there is no Ostrogradsky ghosts.  In fact, in this case, the transformation is not differential, it becomes  algebraic. This class of transformations has been studied in the previous section.

If the theory admits 1 or 2 primary constraints, its canonical analysis is  much subtler and the number of degrees of freedom depends on the number of secondary constraints and on the nature (first or second class) of these constraints.
However, one can easily see that the theory could have up to 5 degrees of freedom, in which case there could be up to two Ostrogradsky ghosts.  In order to avoid such ghosts, it is necessary to eliminate two  degrees of freedom, which is possible only if the transformation~\eqref{derivativetransfo} 
satisfies particular conditions. We will not 
classify these conditions in general (which might be too involved and not so useful) but instead we give a few illustrative examples. 

\subsubsection{Transformations with no higher derivatives of the scalar field}

We first assume that the functions $F_n$ do not depend on $\dot \phi$. In this case the action does not involve second derivatives of the scalar field and the 
phase space is parametrised by 5 pairs of canonical variables  $(g_n,\pi_n; A_{*n},\pi_{*n};\phi,\pi_\phi)$ as there is no need to include in the analysis $A_*$ and its momentum in \eqref{pairofcanon}. A short calculation shows that the momenta 
are given by \eqref{momentadifftransfo},
\begin{eqnarray}
\begin{pmatrix}
\pi_{*1} \\ \pi_{*2}
\end{pmatrix}
=
{}^t J
\begin{pmatrix}
V_{1} \\ V_{2}
\end{pmatrix}
\;\;\; \text{with} \;\; \; 
\begin{pmatrix}
V_{1} \\ V_{2}
\end{pmatrix} = 
J 
\begin{pmatrix}
\dot A_{*1} \\ \dot A_{*2}
\end{pmatrix} 
+ K 
\begin{pmatrix}
A_{*1} \\ A_{*2}
\end{pmatrix} 
+
\dot \phi \begin{pmatrix}
\partial_\phi F_{1} \\ \partial_\phi F_2
\end{pmatrix} 
\end{eqnarray}
where we have introduced the matrices
\begin{eqnarray}
J=\begin{pmatrix}
\partial_{*1} F_1 & \partial_{*2} F_1 \\
\partial_{*1} F_2 & \partial_{*2} F_2
\end{pmatrix} \; , \qquad
K=\begin{pmatrix}
\partial_{1} F_1 & \partial_{2} F_1 \\
\partial_{1} F_2 & \partial_{2} F_2
\end{pmatrix} 
\end{eqnarray}
with the notations 
\begin{eqnarray}
\partial_n F_i= \frac{\partial F_i}{\partial g_n} \, , \qquad \text{and} \quad \partial_{*n} F_i= \frac{\partial F_i}{\partial {A_{*n}}} \, . 
\end{eqnarray}

As a consequence, if $J$ is invertible, the theory does not admit primary constraints.\footnote{To make contact with the previous analysis, the constraint  introduced in \eqref{primarychi} corresponds to $\pi_* \simeq 0$ here and $\chi$~\eqref{chi_diff_noscalar} would be an additional primary constraint.} If $J$ is degenerate but non zero, the theory admits only one primary constraint.\footnote{Or equivalently a second primary constraint if we take into account  the trivial constraint $\pi_* \simeq 0$.} Following the analysis of the previous section, we know that $J$ is degenerate
if the transformation \eqref{derivativetransfo} is such that
\begin{eqnarray}
\label{condnoninvertdiff}
F_2= G(F_1,g_1,g_2,\phi)  \, \quad \text{or} \quad
F_1=G(F_2,g_1,g_2,\phi) \, ,
\end{eqnarray}
where $G$ is an arbitrary function. From these relations, one can easily compute the null vector of the matrix $J$ (or equivalently of its transpose ${}^t J$) in order to obtain
the expression of the primary constraint
\begin{eqnarray}
\chi \; = \; \begin{pmatrix}
\partial_{*2} F_1 & -\partial_{*1} F_1
\end{pmatrix}
\begin{pmatrix}
\pi_{*1} \\ \pi_{*2}
\end{pmatrix} \; = \; \partial_{*2} F_1 \, \pi_{*1}  - \partial_{*1} F_1 \, \pi_{*2} \, \simeq \, 0 \, .
\label{chi_diff_noscalar}
\end{eqnarray}
After that the analysis follows the standard technique. First, one  computes the total Hamiltonian which takes the form
\begin{eqnarray}
H  =  H_0 + \pi_1 A_{*1} + \pi_2 A_{*2} \, + \, \lambda \, \chi , 
\end{eqnarray}
where the Lagrange multiplier $\lambda$ enforces the constraint $\chi \simeq 0$, while the canonical Hamiltonian $H_0$, given by
\begin{eqnarray}
H_0  =   \pi_{*1} \dot A_{*1} + \pi_{*2} \dot A_{*2} + \pi_{\phi} \dot \phi - \frac{1}{2} \left( V_1^2 + V_2^2 + \dot \phi^2 \right) \, ,
\end{eqnarray}
can be expressed  in terms of the phase space variables and does not depend on $\pi_1$ nor $\pi_2$.

Then one  computes the time evolution of $\chi$ which leads to the secondary constraint
\begin{eqnarray}
\psi \; = \; \{ \chi, H\} \; = \;  \{ \chi, H_0\} +  A_{*1} \frac{\partial \chi}{\partial g_1}+  A_{*2}  \frac{\partial \chi}{\partial g_2} - \pi_1  \, \partial_{*2} F_1 + \pi_2 \, \partial_{*1} F_1  \, \simeq \, 0 \;  .
\end{eqnarray} 
To go further, we should compute the stability under time evolution of $\psi$ and keep on doing it until we enumerate all the constraints of the theory, and then  we should classify them into first and second class. However, this is quite a difficult task if we do not specify the transformation explicitly, what we will do later on. Before studying explicit examples, let us discuss the different situations we could encounter in principle.
\begin{itemize}
\item 
The Dirac analysis closes here and the two constraints are second class. In that case, the theory admits 4 degrees of freedom in the 
physical
space. Hence, the transformation is not invertible. There
is one extra degree of freedom which is an Ostrogradsky ghost. Indeed, the constraint $\psi \simeq 0$ enables us to get rid of one of two $\pi_n$ variables, but the remaining one
makes the Hamiltonian unbounded from below. 

\item The Dirac analysis closes here and the two constraints are now assumed be first class. If this happens, the theory would admit 3 degrees of freedom, as the original one. However one of these degrees of freedom
would be an Ostrogradsky ghost. 
This suggests that the transformation is again non-invertible, since one ``regular'' degree of freedom has been 
replaced with
a ghost.\footnote{
An example of such a situation is a transformation
given by
a combination of a non-invertible algebraic transformation and a non-invertible differential transformation. The algebraic transformation would eliminate one regular degree of freedom (as we explained in the previous section) while the differential one would add an Ostrogradsky ghost. In this case, there is no relation between the degree of freedom which has disappeared and the ghost which has been added, therefore the transformation is non-invertible.}

\item The two constraints commute one with the other and requiring the stability under time evolution of $\psi$ leads to (at least) two new constraints. If this is the case, all these constraints would form a set of second class constraints and the number of degrees of freedom would be reduced to (at most) $3$. Even though we have not shown it explicitly, we could expect the theory to have no Ostrogradsky ghost among its degrees of freedom as 
one could make use of the constraints to solve $\pi_n$ in terms of the other phase space variables. The consequence would be that the Hamiltonian is no more linear in these variables. Notice that invertible transformations necessarily fall in this case as they lead to theories with no extra degrees of freedom compared to the original one.
\end{itemize}

\subsubsection{On invertible differential transformations}
Now, we are particularly interested in the case where the transformation is invertible. Such transformations have been recently examined from a different point of view \cite{Babichev:2019twf,Babichev:2021bim}. 
The key observation for the study is that, for the transformation to be invertible, one has to express $g_n$ in terms of $\tilde g_n$ without integrating equations even though the transformation
is differential. 
A priori, it does not seem obvious how to meet this requirement, however it is satisfied if,
\begin{enumerate}
\item First, there exists a function $W(g_1,g_2;\dot g_1,\dot g_2 ;\phi)$ such that the functions $F_n$ themselves can be written in the form
\begin{eqnarray}
\tilde g_n
=
F_n(
g_1,
g_2; \dot {
g}_1, \dot {
g}_2;\phi) = G_n(W,g_1,g_2,\phi) \, , 
\end{eqnarray}
where the functions $G_n$, hence $F_n$, depend on $(\dot g_1,\dot g_2)$ through $W$ only.
\item Second, $W$ satisfies a relation of the form
\begin{eqnarray}
\label{Wtilde}
 \tilde W \, = \,  {I}\left(W \right) \, \quad \text{with} \quad \tilde W=W(\tilde g_1,\tilde g_2; \dot {\tilde g}_1, \dot {\tilde g}_2;\phi) \, ,
\end{eqnarray}
 for a given function $I(z)$. Let us emphasize that $\Tilde{W}$ is exactly the same function as $W$ but evaluated on the variables $\Tilde{g}_n$ and their derivatives instead of $g_n$.
 \end{enumerate}
 One can immediately realise that such a transformation
 is
 invertible (without the need of an integration) in an open set at the vicinity of any point $(g_1,g_2,\dot g_1,\dot g_2)$  when, in addition to the above mentioned two conditions, we also assume $I'(W) \neq 0$. In this case one can invert \eqref{Wtilde} and express $W=\tilde{I}(\tilde W)$. Then, the two differential equations defining the transformation
 \begin{eqnarray}
 \label{transforG}
 \tilde g_n = G_n (\Tilde{I}(\tilde W), g_1,g_2,\phi) 
 \end{eqnarray}
 become in fact algebraic equations between 
 $(g_1,g_2;\tilde g_1,\tilde g_2;\tilde W)$ without an explicit dependence on $\dot{\tilde g}_n$. As a consequence, if the Jacobian of the transformation \eqref{transforG}  for $(g_1,g_2)$
 is non-degenerate, one can express $(g_1,g_2)$ in terms of
$(\tilde g_1,\tilde g_2,\tilde W)$ and the transformation is invertible.
 
 In fact there is a relation between  
 this invertibility property and the canonical analysis we 
 performed in the previous section. 
 Indeed we can show that the action $S[g_1,g_2,\phi]$ propagates the expected three degrees of freedom (with no Ostrogradsky ghosts), 
 because of invertibility,
 which  means that the canonical analysis should lead to a set of 4 secondary constraints. 

\subsubsection{Example: conformal-like transformation}

Let us illustrate this is indeed the case  with a concrete example. As the scalar field does not play any role in the definition of the invertibility,
we assume that there is no scalar field in the theory, for simplicity,
\begin{eqnarray}
S[g_1,g_2
] \; = \;  \frac{1}{2} \int \, \dd t \, \left( {\dot {\tilde g}_1}^2 +  {\dot {\tilde g}_2}^2\right)\, .
\end{eqnarray}
Furthermore, we consider a conformal-like transformation of the variables $(g_1,g_2)$ defined by
\begin{eqnarray}
\label{Example1}
\tilde g_n =
F_n (g_1, g_2 \, ; \, \dot g_1 , \dot g_2 ) \, = \, G(W) \, g_n \qquad \text{with} \qquad W = \dot g_1 \, g_2 - \dot g_2 \, g_1 \, ,
\end{eqnarray}
and $G$ is an arbitrary function at this stage. It is immediate to see that this transformation falls in the class of  differential transformations described in the previous subsection because
\begin{eqnarray}
\tilde W = I(W) \, , \qquad I(z) = z  \,G^2(z)  \, .
\end{eqnarray}
Hence the transformation above \eqref{Example1} is invertible whenever $I'(z)$ is not a vanishing function. 

\medskip

Let us perform the canonical analysis of the action $S[g_1,g_2]$. There are many ways to make the analysis and here, we  follow the method of the previous section
and {we work in the phase space $(g_n,\pi_n;A_{*n},\pi_{*n})$}.
Hence, we first compute the momenta $\pi_{*n}$ associated to the variables $A_{*n}$, 
\begin{eqnarray}
\pi_{*1} = G'(W) \, g_2 (g_1 V_1 + g_2 V_2) \, , \qquad  \pi_{*2} = -G'(W) \, g_1 (g_1 V_1 + g_2 V_2)  \, ,
\end{eqnarray}
where $G'=dG/dW$. The variables $V_n$ were defined in \eqref{defofVn} and are explicitly given by,
\begin{eqnarray}
V_n = G'(W)\, ( \dot A_{*1} g_2 - \dot A_{*2} g_1)g_n + G(W) \, A_{*n} \, .
\end{eqnarray}
As expected, one finds the primary constraint 
\begin{eqnarray}
\chi_{1} \; = \; g_1 \, \pi_{*1} + g_2 \, \pi_{*2} \; \simeq \; 0 \, .
\end{eqnarray}
As a consequence, the total Hamiltonian is given by $H=H_0 + \lambda \chi_1$ where the Lagrange multiplier $\lambda$ imposes the constraint $\chi_1 \simeq 0$ and 
the canonical Hamiltonian $H_0$ is of the form,
\begin{eqnarray}
 H_0 =  \tilde H_0 +  \pi_1 A_{*1} + \pi_2 A_{*2} \, , 
 \end{eqnarray}
 with 
 \begin{eqnarray}
\tilde H_0 = \frac{\bigl(\pi_{*1} - GG'(g_1 A_{*1}  + g_2 A_{*2} )g_2\bigr)^2}{2 G'^2 g_2^2 (g_1^2+g_2^2)} - \frac{1}{2} G^2 (A_{*1}^2+A_{*2}^2) \, .
\end{eqnarray}
Notice that the function $G$ and its derivative $G'$ are evaluated on $W \simeq g_2 A_{*1} - g_1 A_{*2}$. 

In the sequel, it is convenient to introduce the variable $U$ defined by,
\begin{eqnarray}
U = g_1 A_{*1} + g_2 A_{*2} \, ,    
\end{eqnarray} 
and to express the Hamiltonian as a function of $(W,U)$ instead of $(A_{*1},A_{*2})$,
\begin{eqnarray}
\label{tildeH0linear}
 \tilde H_0 \; = \; \frac{1}{2G'^2(g_1^2 + g_2^2) }\left(
 \frac{\pi_{*1}^2}{g_2^2}  - \frac{2 G'G}{g_2}  U \pi_{*1} \right) - \frac{(WG)^2}{2 (g_1^2 + g_2^2)}  \, .
\end{eqnarray}
For the following, it is interesting to underline that $\tilde H_0$, hence $H_0$ itself, is linear in the variable $U$.

Requiring the stability of the primary constraint under time evolution leads to the secondary constraint $\chi_2 \simeq 0$ with 
\begin{eqnarray}
\chi_2 & = & -g_1 \frac{\partial \tilde H_0}{\partial A_{*1}}  -g_2 \frac{\partial \tilde H_0}{\partial A_{*2}} + \pi_{*1} A_{*1} + \pi_{*2} A_{*2} - \pi_1 g_1 - \pi_2 g_2  \nonumber \\
 & = & -(g_1^2 + g_2^2) \frac{\partial \tilde H_0}{\partial U} + \pi_{*1} A_{*1} + \pi_{*2} A_{*2} - \pi_1 g_1 - \pi_2 g_2 \, \nonumber \\
 & \simeq  & \frac{W G' + G }{g_2 G'}  \pi_{*1} - \pi_1 g_1 - \pi_2 g_2 \,  .
\end{eqnarray}
To go further in the canonical analysis, one has to compute the Poisson bracket between the two constraints which is given by
 \begin{eqnarray}
 \{ \chi_1 , \chi_2 \}  =  - 
 2
 \chi_1 + g_1^2 \frac{\partial^2 \tilde H_0}{\partial A_{*1}^2} + 2 g_1 g_2  \frac{\partial^2 \tilde H_0}{\partial A_{*1} \partial A_{*2}} + 
 g_2^2 \frac{\partial^2 \tilde H_0}{\partial A_{*2}^2} 
 = - 
 2
 \chi_1 + (g_1^2+g_2^2)^2 \frac{\partial^2 \tilde H_0}{ \partial U^2} \, .
 \end{eqnarray}
 As a consequence, due to the fact that $\tilde H_0$ is linear in $U$ \eqref{tildeH0linear},  we see that 
 \begin{eqnarray}
 \{ \chi_1 , \chi_2 \} \simeq 0 \, .
 \end{eqnarray}
  Therefore, the canonical
 analysis does not stop here and we have to compute the time evolution of the secondary constraint $\chi_2$. A direct calculation shows that
 \begin{eqnarray}
 \label{chi2dot}
 \dot \chi_2 & = & \{ \chi_2,H_0\} \nonumber \\
 &\simeq &(G + 2W G')  \frac{
 \bigl(g_1 G +(g_1 W - g_2 U)G'\bigr) \pi_{*1} - g_2(g_1^2+g_2^2) G' \pi_1
 +g_2^2WG^2G'
  }{g_2^2 (g_1^2+g_2^2) G'^2}\, ,
 \end{eqnarray}
 where the equality holds up to terms proportional to $\chi_1$ and $\chi_2$. 
 
  First we assume that $\dot \chi_2$ does not trivially vanishes, which means that $G+2W G' \neq 0$. In that case, 
  requiring the stability under time evolution of $\chi_2 \simeq 0$ leads to a new tertiary constraint with,
 \begin{eqnarray}
 \chi_3 \; = \; \bigl(g_1 G +(g_1 W - g_2 U)G'\bigr) \pi_{*1} - g_2(g_1^2+g_2^2) G' \pi_1 
 +g_2^2WG^2G'
 \, .
 \end{eqnarray}
 Again, we study its time evolution and requiring its stability leads generically to a new constraint $\chi_4 \simeq 0$ with
 \begin{equation}
 \chi_4 \simeq 
\pi_{*1}-g_2UGG'
\,.
 \end{equation}
In general, the analysis  stops here with  4 constraints.
We can compute the associated Dirac matrix~$\Delta$ whose coefficients are given by $\Delta_{nm}= \{\chi_n,\chi_m\}$ and whose rank
 gives the number of second class constraints. In the case where $\Delta$ is invertible, the four constraints are second class,
 one can fix {4 out of the 8 variables in the phase space}
 and the theory admits only two degrees of freedom. Furthermore, the constraints enable us to solve $\pi_1$ and $\pi_2$ in terms of the other variables so that 
the Hamiltonian is no longer linear in those variables. Hence, the two degrees of freedom are not Ostrogradsky ghost. As expected, this analysis is consistent with the fact that the transformation is invertible.

\medskip

Now, let us study the case where $G+ 2W G' =0$, which means that
 \begin{eqnarray}
 G = \frac{G_0}{\sqrt{W}} \, ,
 \label{conformalG}
 \end{eqnarray}
 where $G_0$ is a 
 constant.
In this case, $\tilde g_n$ is invariant under the ``conformal transformation'' of $g_n$, that is $g_n(t) \longmapsto \Omega (t)g_n(t)$ for an arbitrary function $\Omega(t)$, as one can see from the definition \eqref{Example1} of $\tilde g_n$.
 Hence, the action possesses the ``conformal'' symmetry.
 Now, requiring the stability under time evolution of the secondary constraint \eqref{chi2dot} does not lead to any new constraint. Therefore, the constraints analysis stops  with a set of two first class constraints~$(\chi_1,\chi_2)$ 
 \begin{eqnarray}
 \chi_1  \; \simeq \; g_1 \, \pi_{*1} + g_2 \, \pi_{*2} \, , \qquad
 \chi_2 \; \simeq \; - \frac{W}{g_2}\pi_{*1}- g_1 \pi_1 - g_2 \pi_2  \, ,
 \end{eqnarray}
 or equivalently $(\chi_1,\tilde\chi_2)$  with
 \begin{eqnarray}
 \tilde \chi_2 \; \simeq \; A_{*1} \pi_{*1} +   A_{*2} \pi_{*2} +   g_1 \, \pi_{1} + g_2 \, \pi_{2} \, .
 \end{eqnarray}
 As a consequence, the theory admits 2 degrees of freedom. The presence of first class constraints is associated to symmetries of the theory and, more precisely, the infinitesimal transformations
 associated with these first class constraints on the variables $g_n$ and $A_{*n}$ are given by
 \begin{equation}
 \begin{aligned}
 \delta_{1} (\epsilon) A_{*n} &= 
\epsilon  \{ \chi_1 , A_{*n} \} = - \epsilon g_{ n} 
\, , \qquad &  \delta_{1} (\epsilon)  g_{n} &= \epsilon \{ \chi_1 , g_{n} \} = 0 \, , \\
 \delta_{2} (\epsilon)  A_{*n} &= \epsilon \{ \tilde \chi_2 , A_{* n} \} = - \epsilon  A_{*n} \, ,\qquad &  \delta_{2}(\epsilon)  g_{n} &= \epsilon \{ \tilde \chi_2 , g_{n} \} = -\epsilon g_n \, ,
\end{aligned}
\label{delta12def}
 \end{equation}
 and similar relations can be computed for the momenta. It is immediate to see that the Lagrangian is invariant under these transformations as
 \begin{eqnarray}
 \delta_{1}(\epsilon)  \left[\frac{g_n}{\sqrt{W}} \right] = - \frac{g_n}{
 2
 W^{3/2}} \delta_{1}(\epsilon) W = 0 \, , \qquad
  \delta_{2}(\epsilon)  \left[\frac{g_n}{\sqrt{W}} \right] = \frac{2 W  \delta_{2}(\epsilon)  g_n - g_n  \delta_{2}(\epsilon)   W}{2 W^{3/2}} = 0 \, .
 \end{eqnarray}
 One recovers the 
 conformal invariance of the original action generated by the constraint\footnote{Let us explain why \eqref{constraint2conformaltrsf} is associated with a conformal transformation $g_n(t) \longmapsto \Omega(t)\, g_n(t)$. Indeed, such a transformation leads to a transformation $\dot g_n \longmapsto \Omega \, \dot g_n + \dot\Omega \,  g_n$\,, which translates into $A_n \longmapsto \Omega \, A_n + \dot\Omega \,  g_n$when reformulated in the phase space. Hence, when one considers  infinitesimal conformal transformations with $\Omega = 1 + \epsilon$\,, the canonical variables transform  according to $\delta g_n = \epsilon \, g_n$ and  $\delta A_n = \epsilon \, A_n + \dot\epsilon \, g_n$\,. Therefore, we see that \eqref{constraint2conformaltrsf} is nothing but this infinitesimal transformation expressed in terms $\delta_1(\epsilon)$ and $\delta_2(\epsilon)$ given by~\eqref{delta12def}. }
\begin{eqnarray}
\delta_c(\epsilon) = - \delta_1(\dot \epsilon) - \delta_2 (\epsilon) \, .
\label{constraint2conformaltrsf}
\end{eqnarray}
The second transformation, associated with $\delta_1(\epsilon)$, corresponds to an invariance of $W$ which is easily seen\footnote{Indeed, we have $\delta_1(\epsilon) W= g_2 (-\epsilon g_1) - g_1 (-\epsilon g_2)=0$.} from the expression $W \simeq g_2 A_{*1}-g_1 A_{*2}$\,.
 
 \medskip
 
 However, the two constraints do not enable us to solve the momenta $\pi_1$ and $\pi_2$ in terms of the remaining variables. Hence, the Hamiltonian is still unbounded from below and one of these two degrees of freedom is certainly an Ostrogradsky ghost. To see this is indeed the case, we 
 reformulate
 the action in terms of the variable $q=g_1/g_2$. A direct calculation shows that the resulting action is given by
 \begin{eqnarray}
 {S[q]} 
 = \frac{
 G_0^2
 }{8}\int \dd t \, \frac{1}{\dot q^3} \left[ (1+ q^2) \ddot q^2 + 4 \dot q^4 - 
 4
 q \dot q^2 \ddot q \right] \, ,
 \end{eqnarray}
 which makes clear that the theory propagates two degrees of freedom, one of them being an Ostrogradsky ghost.

\subsubsection{Non-invertible transformations with higher derivatives of the scalar field}
\label{sec:toymodel_noninv-diff-trsf}
The aim of this section is to study further the case of non-invertible transformations. We still consider the conformal invariant model but we generalise the transformation \eqref{Example1} to the case where
\begin{eqnarray}
\label{generalWalphabeta}
W \; = \; \alpha \, (\dot g_1 \, g_2  - \dot g_2 \, g_1) \, + \, \beta \, g_1 g_2 \, .
\end{eqnarray}
Furthermore, we assume that $\alpha$ and $\beta$ are not constant but depend on a scalar field $\phi$ and its derivatives.  
If $G$ is given by~\eqref{conformalG},
the resulting theory is still conformal invariant (which is associated to the non-invertibility of the  transformation, as in mimetic gravity) and we ask the question of the number and the nature of the degrees of freedom. For that, we proceed as in the previous
case, we introduce 
the  
variable 
$q =g_1/g_2$,
we reformulate the variables $\tilde g_n$ in terms of $q$ as follows,\footnote{We set $G_0=1$ without loss of generality with a simple rescaling of $\alpha$ and $\beta$.
Also we assume $g_2>0$ for simplicity.}
\begin{eqnarray}
\tilde g_1 = 
q
\left(\alpha \dot q + \beta q \right)^{-1/2} \, , \qquad
\tilde g_2 =  \left(\alpha \dot q + \beta q \right)^{-1/2} \, ,
\end{eqnarray}
 and then we compute the 
 action $S[q,\phi] $ whose expression is cumbersome and not necessary for our purposes.
If $\alpha$ and $\beta$ depend on $\phi$ and its first derivative, the Lagrangian involves second derivatives of $\phi$ and $q$, and then the theory could propagate up to 4 degrees of freedom, two of them being Ostrogradsky ghost. Let us
illustrate this aspect with few examples.

\medskip

We start with the case of an algebraic transformation with
$\alpha=0$, which corresponds to a toy-model for the original mimetic action. In that case, it is convenient to change variables and to introduce
\begin{eqnarray}
s = \sqrt{q} \, , \qquad \gamma = 1/\sqrt{\beta} \, ,
\end{eqnarray}
where $q$ and $\beta$ are supposed to be positive,
so that the action takes the form
\begin{eqnarray}
S[s,\phi] = \frac{1}{2} \int \dd t \, \left[ \left(s^2+\frac{1}{s^2}\right) \dot \gamma^2 + \gamma^2 \left(1+\frac{1}{s^4}\right) \dot s^2 + 2 \left(1-\frac{1}{s^4}\right) \gamma s \dot \gamma \dot s \right] \, .
\end{eqnarray}
If $\beta$ depends on $\phi$ only, the Lagrangian depends on $r$, $\phi$  and its first derivatives only, then the theory propagates 2 degrees of freedom, none of them being Ostrogradsky ghosts. If $\beta$ depends on $\dot \phi$ (as it is the case in mimetic gravity), the Lagrangian involves second derivatives of the scalar field. To study the numbers and the nature of the degrees of freedom,
we compute the kinetic Lagrangian
\begin{eqnarray}
L_{\textrm{kin}} = \frac{1}{2} \left[ 
s^2 \left( 1 + \frac{1}{s^4}\right) \gamma_{\dot \phi}^2 \ddot \phi^2 + \gamma^2  \left( 1 + \frac{1}{s^4}\right) \dot s^2 + 
2  \left( 1 - \frac{1}{s^4}\right) \gamma \gamma_{\dot \phi} s \ddot \phi \dot s 
\right] \, ,
\end{eqnarray}
where $\gamma_{\dot \phi}$ is the derivative of $\gamma$ with respect to $\dot \phi$.  Then, we easily extract the kinetic matrix $K$ and its determinant
\begin{eqnarray}
K =\frac{1}{s^4} \begin{pmatrix}
s^2 \left( 1 + {s^4}\right) \gamma_{\dot \phi}^2 & \left( s^4 - {1}\right) \gamma \gamma_{\dot \phi} s \\
\left( s^4 - {1} \right) \gamma \gamma_{\dot \phi} s  &  \gamma^2  \left( 1 + {s^4}\right) 
\end{pmatrix} \, , \qquad
\det K = \frac{4 }{s^2}   \gamma^2 \gamma_{\dot \phi}^2 \, .
\end{eqnarray} 
As $K$ is non-degenerate, the theory propagates 3 degrees of freedom and one of them is a ghost.

\medskip

The general case
of differential transformations with non-vanishing
 $\alpha$ and $\beta$ is studied in the appendix \ref{AppA}. We show that
the extended mimetic theory discussed in section~\ref{sec:extended-mimetic} corresponds to the case where the two functions are 
given by
$\alpha = \alpha(q, \phi,\dot\phi)$ and $ \beta = \beta(q,\phi,\dot\phi, \ddot\phi)$. The theory could propagate up to two Ostrogradsky ghosts in this case.

\subsection{Summary: Invertible vs. non-invertible transformations}
Before generalising the study of this section to higher derivative scalar-tensor theories, it might be useful to make a quick summary at this stage. 
We considered a simple toy model which describes two degrees of freedom mimicking two polarisation of a graviton 
and one free scalar-like degree of freedom \eqref{freeaction}. Then we have introduced two types of transformations, the algebraic ones and the differential ones, which enable us to transform the free theory to a more complex theory. In both cases, we gave conditions for these transformations to be invertible.

We show that, as expected, when these invertibility conditions are satisfied, the number of degrees of freedom remain unchanged and the transformed theory does not propagate any Ostrogradsky ghosts, even though the transformation involves derivatives of the scalar field and the new action involves higher derivatives of the scalar variable $\phi$.

The case of a non-invertible transformation is much subtler. When the transformation is algebraic, we could make a general study. In particular, we showed that, when the transformation does not involve time derivatives of the scalar-like degree of freedom $\phi$, the new theory propagates only two degrees of freedom with no Ostrogradsky ghost, i.e.\ the non-invertible transformation kills one degree of freedom. When the transformation involves $\dot\phi$, the new theory admits three degrees of freedom, one of them being an Ostrogradsky ghost. In other words, the non-invertibility of the transformations kills one degree of freedom and, at the same time, the presence of $\dot \phi$ in the transformation awakes  an Ostrogradsky ghost. 

When the transformation is differential, the analysis is more complex. We have not made a general study of this case but instead we have introduced and studied in details an interesting example with a conformal-like non-invertible transformation. In this example, the non-invertibility leads to a new theory which always propagates at least one Ostrogradsky ghosts: when the transformation does not involve $\dot \phi$, there is only one Ostrogradsky ghost while there are several such ghosts when the transformation does  involve time derivatives of $\phi$.

\section{Generalised Transformations of the metric}
\label{sec:gentrsf-metric}

In this section, we adapt and extend the previous analysis to scalar-tensor theories. We start by recalling known properties on  generalised transformations of the metric which involve higher derivatives of the scalar field and/or the metric. Then, we introduce a new class of generalised conformal transformations and we study their properties.

\subsection{From disformal to  higher derivative transformations of the metric}
We consider higher derivative transformations of the metric of the form,
\begin{eqnarray}
\label{GenTran}
g_{\mu\nu} \longmapsto \tilde{g}_{\mu\nu}(g_{\alpha \beta}, R_{\alpha\beta\rho\sigma}\, ;\phi,\nabla_\alpha\phi,\nabla_\alpha\nabla_\beta \phi; \cdots) \, ,
\end{eqnarray}
where the new metric $\tilde{g}_{\mu\nu}$ depends on the derivatives of $g_{\mu\nu}$ through the Riemann tensor and the covariant derivatives of $\phi$. We will use the standard notations $\phi_\mu=\nabla_\mu \phi$ and $\phi_{\mu\nu}=\nabla_\mu \nabla_\nu \phi$ for the 
covariant
derivatives of the scalar field. 
Generically, without any assumptions, such a transformation is not invertible as one needs to integrate $g_{\mu\nu}$ to get its expression in terms of $\tilde{g}_{\mu\nu}$ and the scalar field. 

However, as in the previous section, we will give (sufficient) conditions for the transformation to be invertible. For that, it is useful to proceed as in the case of the toy models and to distinguish  between algebraic and differential transformations. 

\subsubsection{Algebraic transformations of the metric: standard disformal transformations}
The usual disformal transformations of the form,
\begin{eqnarray}
    g_{\mu\nu} \longmapsto \tilde{g}_{\mu\nu} = A(\phi,X) g_{\mu\nu} + B(\phi,X) \phi_\mu \phi_\nu \, ,
\end{eqnarray}
where $A(\phi,X)$ and $B(\phi,X)$ are a priori arbitrary functions of $\phi$ and $X=\phi_\mu \phi^\mu$, are algebraic transformations of the metric as they do not involve any derivatives of the metric.

A necessary and sufficient condition for the transformation to be invertible is that the Jacobian matrix given by,
\begin{eqnarray}
\label{Jacob2}
J_{\mu\nu}^{\alpha \beta} = \frac{\partial \tilde{g}_{\mu\nu} }{\partial g_{\alpha \beta}} \, ,
\end{eqnarray}
to be invertible. A direct calculation shows that this is  satisfied when
\begin{eqnarray}
    X^2 B_X - A + XA_X \neq 0 \, ,
\end{eqnarray}
where $F_X$ denotes the derivative of any function $F$ with respect to $X$. If this is not the case, the transformation is non-invertible and one cannot express $g_{\mu\nu}$  in terms of $\tilde{g}_{\mu\nu}$ in a unique way. This is very well-known and details can be found in the literature (see e.g.\ \cite{Zumalac_rregui_2014}). 

\subsubsection{Differential transformations of the metric: examples}
Differential transformations of the metric have been much less studied. Very recently \cite{Takahashi_2022,Takahashi_2023}, interesting examples of such transformations, which involve derivatives of the metric, have  been considered, mostly in the context of scalar-tensor theories. In those examples, the metric is transformed as follows
\begin{eqnarray}
\label{GenDisTransf}
g_{\mu\nu} \; \longmapsto \; \tilde{g}_{\mu\nu} = F_0 g_{\mu\nu} + F_1 \phi_\mu \phi_\nu + 2 F_2 \phi_{(\mu} X_{\nu)} + F_3 X_\mu X_\nu \, ,
\end{eqnarray}
where $F_i$ are functions of the scalar field $\phi$ and also $X$ , $Y$, $Z$ where the definition of $X$ has been recalled above while the last two variables are defined by
\begin{eqnarray}
\label{defYZ}
Y=\phi^\mu X_\mu \, , \quad Z= X^\mu X_\mu \, .
\end{eqnarray}
Such transformations involve the first derivatives of the metric through the derivatives of $X$, which are given by
\begin{eqnarray}
X_\mu = 2 \phi^\nu \, \nabla_\nu \phi_\mu = 2 \phi^\nu \, \left( \partial_\mu \partial_\nu \phi - \Gamma_{\mu\nu}^\lambda \phi_\lambda \right) \, ,
\end{eqnarray}
because the Christoffel symbols $\Gamma_{\mu\nu}^\lambda$ are obviously expressed in terms of the first partial derivatives of $g_{\mu\nu}$. As a consequence, in general (without specific assumptions), the transformation \eqref{GenDisTransf} is not invertible. 

In order to illustrate this fact, following \cite{Takahashi_2022,Takahashi_2023}, it is instructive to compute the inverse metric 
$\tilde{g}^{\mu\nu}$ from which we get $\tilde X$,
\begin{eqnarray}
\label{XtildeX}
\tilde X = \tilde g^{\mu\nu} \phi_\mu \phi_\nu = Q(\phi,X,Y,Z) \, ,
\end{eqnarray}
where $Q$ can be easily expressed in terms of the functions $F_i$. Hence, in general, $\tilde X$ depends on $X$ and  its derivatives $X_\mu$, and then one needs to integrate a partial differential equation to obtain its expression  in terms of $\tilde X$ for instance. If such an integration is needed, the transformation between $X$ and $\tilde X$ is clearly not invertible.

The conditions for these transformations to be invertible have been well studied in \cite{Takahashi_2022,Takahashi_2023}. It has been realised that the key assumption is that the relation between $X$ and $\tilde X$ is algebraic in the sense that 
$\tilde X$ depends on $\phi$ and (non-trivially on) $X$ only which means that
\begin{eqnarray}
\label{condTak}
\frac{\partial Q}{\partial X} \neq 0 
\qquad \text{and} \qquad \frac{\partial Q}{\partial Y} =  \frac{\partial Q}{\partial Z} =  0 \, .
\end{eqnarray}
If this is indeed the case,
one does not need anymore to integrate an equation to express $X$ in terms of $\tilde X$, one only needs to solve an algebraic equation (which can be done locally when $Q$ satisfies the hypothesis needed to apply the implicit function theorem) and get
\begin{eqnarray}
\label{Xftilde}
X = \tilde Q\bigl(\phi, \tilde X\bigr) \, ,
\end{eqnarray}
where $\tilde Q$ is the inverse of $Q$ for the composition law. Furthermore, as shown in \cite{Takahashi_2022,Takahashi_2023}, the relations between the variables ($\tilde Y, \tilde Z)$ and $(Y,Z)$ is algebraic, hence one can express the former in terms of the later if the corresponding Jacobian\footnote{The Jacobian is the following two dimensional matrix   
\begin{eqnarray}
\left(
\begin{array}{cc}
   \partial \tilde Y/\partial Y  & \partial \tilde Y/\partial Z  \\
    \partial \tilde Z/\partial Y & \partial \tilde Z/\partial Z
\end{array}
\right) \,.
\end{eqnarray} 
} is invertible.

As a consequence, one can substitute the expressions of $X$ \eqref{Xftilde} and also $(Y,Z)$ into the transformation \eqref{GenDisTransf} so that
$\tilde g_{\mu\nu}$ does not depend on the derivatives of $g_{\mu\nu}$ anymore as they have been absorbed into $\tilde X$. Hence, the relation \eqref{GenDisTransf} becomes algebraic 
for $(g_{\mu\nu};\tilde g_{\mu\nu}, \tilde X, \phi)$
and then there is no need to integrate any equation to invert the relation. Nonetheless, one has to be sure that the remaining algebraic relation between $g_{\mu\nu}$ and $\tilde{g}_{\mu\nu}$ is invertible, which is easy to verify with the Jacobian matrix, similarly to what we did above for disformal transformations \eqref{Jacob2}. 

\subsection{Invertible differential transformations of the metric}
Now, we extend the  construction presented in the previous subsection to introduce a new class of invertible differential transformations of the metric.

\subsubsection{From the previous example to new classes of transformations}
What are the lessons to be drawn from the example of the previous subsection? We clearly see  that one way to make a transformation of the form \eqref{GenTran} 
invertible consists in requiring that the transformation satisfies the following  properties.

First, we assume that there exists a set of $N$ scalars $C_n[g,\phi]$ (where $1 \leq n \leq N$) that can be constructed from the metric, the scalar field and their derivatives, which transform in an algebraic way under the transformation  \eqref{GenTran}. More precisely, we assume that the transformed scalars $\tilde C_n=C_n[\tilde g,\phi]$ are related to the non-transformed ones by algebraic relations of the form
\begin{eqnarray}
\label{condCn}
\tilde C_n = Q_n (C_1,\cdots,C_N) \, ,
\end{eqnarray}
where $Q_n$ are arbitrary functions at this stage. Furthermore, we assume that these algebraic transformations are (locally) invertible, i.e.\ one can express the scalars $C_n$ in terms of $\tilde C_n$.

Second, we require that all derivatives of the metric enter  in the transformation \eqref{GenTran} through first-order partial derivatives of the scalar $C_n$ only, i.e.\
\begin{eqnarray}
\label{TransfowithCn}
g_{\mu\nu} \longmapsto \tilde{g}_{\mu\nu}(g_{\alpha \beta}
, C_n
; \partial_\alpha C_n) \, .
\end{eqnarray}
Third, when we express the transformation as \eqref{TransfowithCn}, we assume the following Jacobian matrix,
\begin{eqnarray}
J_{\mu\nu}^{\alpha \beta} = \frac{\partial \tilde{g}_{\mu\nu} }{\partial g_{\alpha \beta}} \, ,
\end{eqnarray}
to be invertible.
We regard $C_n$ as fixed variables when we evaluate the Jacobian matrix.

Hence, it is easy to see that these  conditions  are sufficient to insure the invertibility of the  general transformations \eqref{GenTran}. This is exactly what happens in the previous example 
which involves the scalars $\phi,X,Y,Z$ which obey algebraic relations \eqref{Xftilde}.
A more general transformation would be of the form,
\begin{eqnarray}
    \Tilde{g}_{\mu\nu} \; =\; \Omega(
    C_n) \, g_{\mu\nu} 
    + \sum_{p,q} D_{p,q}(
    C_n) \, \partial_{(\mu} C_{p} \partial_{\nu)} C_q \, ,
\end{eqnarray}
where we use the notation $(\mu,\nu)$ for the symmetrisation, while $\Omega$
and 
$D_{p,q}$ are functions of 
$C_n$.\footnote{
$\phi$ itself is included as one of the scalars $C_n$.}

\medskip

To summarise, we have found sufficient conditions  for a differential transformation to be invertible. Let us underline that it is not clear whether or not the conditions we found are necessary. 
These conditions rely on  the existence of scalars $C_n$ which transform algebraically \eqref{condCn}. However, we have not described any
concrete way to show the existence of such scalars given a transformation. It is in fact the difficult part to construct explicitly these scalars which relies on the calculation of the inverse metric $\Tilde{g}^{\mu\nu}$.
In the following section, we will construct explicit transformations which satisfy  such conditions and we provide an algorithm to construct such scalars.

\subsubsection{Higher order conformal transformations}
\label{sec:higherorderconf}

In this section, we construct a new class of higher order conformal transformations\footnote{Higher derivative conformal transformations have been considered in \cite{Takahashi:2022mew,Tahara:2023pyg} using the Weyl tensor which is well-known to be conformal invariant.} 
\begin{eqnarray}
\label{confmetric}
    \tilde g_{\mu\nu} \; = \; \Omega \, g_{\mu\nu} \, ,
\end{eqnarray}
which satisfy the properties described in the previous subsection, in particular it is possible to construct explicit scalars that satisfy a simple version of \eqref{condCn}. 

For that, it is   convenient to define the  metric $\hat{g}_{\mu\nu}$  by\footnote{We may want to define $\hat g_{\mu\nu}$ with an absolute value of $X$  so that the signatures of the two conformally related metrics are the same whatever the sign of $X$ is.}
\begin{eqnarray}
\label{mimeticmetric}
\hat g_{\mu\nu} \; = \; X \, g_{\mu\nu} \; = \; \phi_\alpha \phi_\beta \, g^{\alpha\beta} \, g_{\mu\nu} \, .
\end{eqnarray}
For obvious reasons, we call $\hat{g}_{\mu\nu}$ the mimetic metric associated to $g_{\mu\nu}$, and it is invariant under any conformal transformation of the metric $g_{\mu\nu}$. 
Hence, the two conformally related metrics $g_{\mu\nu}$ and $\tilde g_{\mu\nu}$ admit the same mimetic metric.

An immediate consequence  is that all tensors constructed from the mimetic metric $\hat{g}_{\mu\nu}$ are invariant under any conformal transformations of the metric $g_{\mu\nu}$. For instance, we can easily construct conformal invariant scalars which involve  first order  derivatives of the metric only as follows,
\begin{eqnarray}
\label{defofCalpha}
C_1=\hat{\phi}_\mu^\mu  \, , \qquad
C_2 = \hat{\phi}_{\mu}^{\nu} \, {\hat{\phi}}^{\mu}_{\nu} \, , \qquad
C_3 = \hat{\phi}_{\mu}^{\nu} \,  {\hat{\phi}}_{\nu}^{\rho}  \, {\hat{\phi}}_{\rho}^{\mu} \, ,
\end{eqnarray}
where we have introduced the notations,
\begin{eqnarray}
\hat g^{\mu\nu} = \frac{1}{X} g^{\mu\nu} \, , \quad \hat{\phi}^\mu = \hat{g}^{\mu\nu}\phi_\nu \, , \quad    \hat{\phi}_\mu^\nu \; = \; \hat{g}^{\nu \rho} \, \hat{\nabla}_\rho \hat{\nabla}_\mu \, \phi\, ,
\end{eqnarray}
with $\hat{\nabla}$ being the covariant derivative compatible with the mimetic metric.
In fact, it is easy to see that any conformal invariant scalars, which involve  first order  derivatives of the metric only, is a function of the three quantities $C_n$ only. Indeed, 
all  scalars that are constructed from contractions that involve
$\hat{\phi}^\nu \hat{\phi}_{\mu\nu}$
vanish because
\begin{eqnarray}
    \hat{\phi}^\nu \hat{\phi}_{\mu\nu} = \frac{1}{2} \partial_\mu \hat{X} = 0 \, ,
\end{eqnarray}
as an immediate consequence of $\hat X\equiv \hat g^{\mu\nu} \phi_\mu \phi_\nu = 1$.
{Also, the Cayley-Hamilton theorem guarantees that $C_{n>3}$ can be expressed using only $C_{n}$ for $n \leq 3$.}
Notice that, one can easily generalise the constructions to scalars that involve higher derivatives with contractions of 
$\hat{\phi}_{\mu\nu\rho}$ for instance.

\medskip

For concreteness, let us give an explicit expression of the scalar $C_1$ in terms of the metric $g_{\mu\nu}$.  For that, we first recall the following property
\begin{eqnarray}
\tilde \phi_{\mu\nu} = \phi_{\mu\nu} - \frac{1}{2\Omega} \left(\Omega_\mu \phi_\nu + \Omega_\nu \phi_\mu - \Omega_\lambda \phi^\lambda \, g_{\mu\nu}\right) \, ,
\end{eqnarray}
which relates  second derivatives of any scalar field associated with $g_{\mu\nu}$ and those associated with $\tilde{g}_{\mu\nu}$ \eqref{confmetric}. When we apply it to $\hat{g}_{\mu\nu}$ (i.e.\ when $\Omega=X$), we obtain
\begin{eqnarray}
\hat{\phi}_{\mu\nu} & = & \phi_{\mu\nu} - \frac{1}{2 X} \left(X_\mu \phi_\nu + X_\nu \phi_\mu - \phi^\lambda X_\lambda \, g_{\mu\nu}\right) \, .  
\end{eqnarray}
Hence, we can easily obtain the  expressions of the scalar $C_n$. For simplicity, we only give $C_1$\footnote{To the best of the authors' knowledge, invertible conformal and disformal transformations whose coefficients depend on $C_1$ were proposed first in Appendix D of~\cite{Domenech:2019syf}.}
\begin{eqnarray}
\label{explicitC1}
C_1= \frac{1}{X}\left(\Box\phi + \frac{1}{ X} {\phi_\lambda X^\lambda}\right) \, .
\label{C1expression}
\end{eqnarray}
In Appendix~\ref{App:derive_C1}, we give an alternative derivation of the invertible higher order conformal transformations depending on $C_1$.

\medskip

Now, it becomes straightforward to construct invertible higher derivative  conformal transformations of the metric. Indeed, 
any conformal transformation of the form,
\begin{eqnarray}
\tilde g_{\mu\nu} \; = \; 
{\Omega(
X,C_n)}
\, g_{\mu\nu} \, , 
\label{invertibleconf}
\end{eqnarray}
can be made invertible with the only requirement that  the relation between $X$ and $\tilde X$,
\begin{eqnarray}
\tilde X = \frac{X}{\Omega (
X,C_n)} \, ,
\label{tildeXeq}
\end{eqnarray}
is invertible, i.e.\ $\partial_X \tilde X \neq 0$ so that one can express (at least locally) $X$ as a function of $\tilde X$ and $\tilde C_n$. On the contrary, the transformation is not invertible only in the case where $\Omega$ is linear in $X$, i.e.\
\begin{eqnarray}
\label{noninvertconf}
\Omega (
X,C_n) = 
-
X\, \Lambda (
C_n) \, ,
\end{eqnarray}
for a given function $\Lambda$.
We will shortly discuss this particular class of transformations later on when we introduce extended mimetic theories, as mimetic theories are constructed from non-invertible metric transformations. 

\subsubsection{Generalisation to disformal transformations}
The  construction of invertible higher derivative transformations can be  extended to general disformal transformations. The idea is simple and consists in combining a higher derivative conformal transformation to a usual disformal transformation. Therefore, we consider a transformation that maps $g_{\mu\nu}$ to $\tilde g_{\mu\nu}$ of the form
\begin{eqnarray}
\label{generaltransfo2}
\tilde g_{\mu\nu} = \Omega \, f_{\mu\nu} \, , \qquad f_{\mu\nu} =  g_{\mu\nu} + B \, \phi_\mu \phi_\nu \, ,
\end{eqnarray}
where $B$ and $\Omega$ depend on $g_{\mu\nu}$, $\phi$ and their derivatives.  We assume  $f_{\mu\nu}$ to be  invertible and its inverse  $f^{\mu\nu}$ given by,
\begin{eqnarray}
f^{\mu\nu} = g^{\mu\nu} - \frac{B}{1+XB} \phi^\mu \phi^\nu \, .
\end{eqnarray}
Therefore, we can generalise the definition of the mimetic metric and we introduce the extended mimetic metric defined by,
\begin{eqnarray}
\hat g_{\mu\nu} = \phi_\alpha \phi_\beta \,  f^{\alpha\beta} \, f_{\mu\nu} \, = \, \frac{ X }{1+XB} (g_{\mu\nu} + B \, \phi_\mu \phi_\nu )  \,= \,\phi_\alpha \phi_\beta \, \tilde{g}^{\alpha\beta} \, \tilde g_{\mu\nu} \, .
\end{eqnarray}
Thus, we proceed as in the previous case and we consider the scalars $C_n$ constructed from $\hat g_{\mu\nu}$. These scalars can be expressed equivalently\footnote{Notice that, contrary to what happens in the case of a higher derivative conformal transformation, we do not have anymore the property that $C_n$ and $\Tilde{C}_n$ are the same scalars evaluated with the two metrics $g_{\mu\nu}$ and $\tilde{g}_{\mu\nu}$ respectively. Instead, they are the same scalars evaluated with $f_{\mu\nu}$ and $\tilde{g}_{\mu\nu}$.} in terms of $g_{\mu\nu}$ or $\tilde{g}_{\mu\nu}$ (through the metric $f_{\mu\nu}$), and then  we assume again that $\Omega$ and $B$ depend only on the variables $(
X,C_n)$. 

The last step consists in requiring that $\tilde X$, which is given by
\begin{eqnarray}
\tilde X = \frac{X}{(1+ X B) \Omega} \, ,
\end{eqnarray}
to be invertible in the sense that one can express $X$ in terms of $\tilde X$ (at least locally).\footnote{Notice that  the transformation is non-invertible for $\Omega$ of the form
\begin{eqnarray}
\Omega(
X,C_n) = \frac{-X}{1 + X B} \,\Lambda(
C_n) \, ,
\end{eqnarray}
where $\Lambda$ depends arbitrarily on $C_n$ and its {first} derivatives.
It can be shown that such a condition coincides with the condition for the full disformal transformation \eqref{generaltransfo2} to be non-invertible.}

\medskip

To conclude this section, let us remark that the class of invertible higher derivative transformations can be extended even more by considering in \eqref{generaltransfo2} an intermediate metric $f_{\mu\nu}$ of the form \eqref{GenDisTransf}
where $F_0=1$ while the functions $F_n$ ($n>1$) depend on $C_n$, their derivatives and also on $X,Y,Z$ \eqref{defYZ}. If we  keep the conditions  \eqref{condTak}, the transformation is still invertible.

\subsection{Extended Mimetic Theories}
\label{sec:extended-mimetic}

In this last section, we make use of the non-invertible transformations of the metric introduced above in \eqref{noninvertconf} 
\begin{eqnarray}
\tilde g_{\mu\nu} \; = \; \Omega(X,C_n) \, g_{\mu\nu} \,  = \, -X \, \Lambda(C_n) \, g_{\mu\nu} \, ,
\label{extmimeticgdef}
\end{eqnarray}
to define a new class of higher order scalar-tensor theories which generalises mimetic gravity theories~\cite{Chamseddine:2013kea,Chamseddine:2014vna}\footnote{These theories are different from the ``extended mimetic gravity" discussed in~\cite{Takahashi:2017pje}, which were defined by applying the conformal transformation $\tilde g_{\mu\nu} = -X g_{\mu\nu}$ to the Horndeski theory.
See also~\cite{Domenech:2023ryc} for a recent attempt to construct new theories using the symmetry under the disformal transformations. } (here, it is convenient to distinguish $X$ from the other scalars $C_n$ in the arguments of the function $\Omega$).
Indeed, as it has been already explained, such a transformation is not invertible because of the conformal invariance of $\tilde g_{\mu\nu} $ and this has nothing to do with the presence of higher derivatives in the transformation. In that sense, it is an extension of the mimetic metric \eqref{mimeticmetric} and we introduce the extented mimetic theory governed by the action 
\begin{eqnarray}
S[g_{\mu\nu},\phi] = S_{EH}[\tilde g_{\mu\nu}] \, .
\end{eqnarray}
We are going to study some aspects of this theory in the simplest case where $\Lambda$ depends on $C_1$ \eqref{defofCalpha} only (and not on its derivative $\partial_\alpha C_1$). Making its full Hamiltonian analysis is rather cumbersome and, for simplicity, we limit our analysis to the case of cosmology where the metric is a flat 
FLRW spacetime and the scalar field depends on $t$ only, i.e.\ 
\begin{equation}
d\tilde s^2 \; = \; \tilde g_{\mu\nu} dx^\mu dx^\nu = -\tilde N^2 dt^2 + \tilde a^2 d\mathbf{x}^2\, ,
\quad
\phi = \phi(t)~\,.
\label{FRW}
\end{equation}
The details of the analysis are given in the Appendix \ref{AppB} where we show that the theory propagates two physical degrees of freedom.

Interestingly, there is one more degree of freedom than in the standard mimetic theory (whose Hamiltonian analysis in an FLRW spacetime has also been recalled in the appendix \ref{AppB}). This extra degree of freedom is due to the presence of higher derivatives of the metric and the scalar field in the transformation \eqref{noninvertconf}. We also notice that the
Hamiltonian constraint is linear in two momenta which might be a sign of the presence of Ostrogradsky ghosts among its degrees of freedom. This last point deserves a careful study that we postpone in a future work. 

\section{Conclusion}
\label{sec:conclusion}

In this work, we studied properties of field transformations involving (higher) derivatives of the fields, motivated by the extension of the disformal transformations, 
which was recently proposed in \cite{Takahashi_2022,Takahashi_2023}.

For this purpose, we first introduced a simple toy model in classical mechanics that mimics scalar-tensor theories and we studied 
higher derivative transformations in such models. More particularly, we studied  how the (non-)invertibility of the transformations 
affects the number and the nature of  the dynamical degrees of freedom, and possible appearance of Ostrogradsky ghosts.

Being in the context of classical mechanics, we introduced two classes of transformations, the algebraic and the differential transformations. 
The former do not involve derivatives of the transformed variables,
while the latter does. We found conditions for these transformations to be invertible. 

We presented examples illustrating that any invertible transformation does not change the number of degrees of freedom and no Ostrogradsky ghosts appear, as it should be. 
We studied non-invertible algebraic transformations and 
we exhibited a subclass of transformations which do not introduce Ostrogradsky ghost, and moreover 
eliminate a degree of freedom. Outside this subclass, the non-invertibility leads to Ostrogradski ghosts.
Concerning non-invertible differential transformations, in general, they introduce extra ghost degrees of freedom.
We have not studied differential transformations in full generality, 
we instead focused 
mainly on what we called conformal-like transformations, which mimic conformal transformations in scalar-tensor theories.  
When such a conformal-like transformation is not invertible, 
we found that at least one Ostrogradsky ghost appears in the theory. 

\medskip

The analysis of these toy models enabled us to better understand higher derivative transformations of the metric in the context of scalar-tensor theories. Indeed, we provided general conditions for the transformation to be invertible and we illustrated these conditions in the particular case of conformal transformations. 
As a result, we found a new class of higher derivative invertible conformal transformations of the metric. 
The construction we proposed is based on the concept of the mimetic metric.

We made use of these higher derivative conformal transformations to
 define extended mimetic theories of gravity, starting from 
 the Einstein-Hilbert action.
It should be stressed that the non-invertibility of the extended conformal transformation constructed in this paper is not due to the presence of higher derivatives but 
is a consequence of a conformal symmetry, as it is the case in the standard mimetic gravity~\cite{Chamseddine:2013kea}. 
We studied the properties of the extended mimetic theories in case of homogeneous FLRW cosmology,
and we found indications that such theories propagate Ostrogradsky ghosts.
However, to make a conclusive statement on the existence of the Ostrogradsky ghosts in the extended mimetic theories, we would need to perform the full Hamiltonian analysis without specifying the background spacetime. We defer such a full-fledged treatment for future works.

\medskip

Let us 
conclude with two 
remarks. The first one concerns the coupling to matter fields. Indeed, it is known that coupling to matter in a higher-derivative theory may lead to instabilities \cite{Deffayet:2020ypa,Naruko:2022vuh,Takahashi:2022mew,Takahashi:2022ctx,Ikeda:2023ntu}. Hence, it 
will be interesting to study this problem in future  within the theories generated by (non-)invertible transformations with higher derivatives. 

The second remark concerns 
possible further generalisation of the transformations of the metric we have studied in this paper.
Indeed, our construction of higher derivative conformal transformations relies on the existence of conformal invariant scalars. 
A natural question is then to ask whether it is possible to use scalars which are not necessarily invariant and satisfy the more general relation~\eqref{condCn}. 
Alternatively, one can also consider transformations of the form
\begin{eqnarray}
\tilde g_{\mu\nu} =   A \, g_{\mu\nu} + B \, \phi_\mu \phi_\nu + C \, \phi_{\mu\nu} + D \, \phi^2_{\mu\nu} + \cdots  \, ,
\end{eqnarray}
where $A$, $B$, $C$, $D$ are functions of scalars constructed from the metric and the scalar field. Such metrics appear when one considers linear perturbations about static and spherically symmetric black holes in DHOST theories \cite{Tomikawa:2021pca,Langlois:2021aji,Langlois:2022ulw}. 
It would be interesting to establish 
conditions for them to be invertible, which 
would help to understand better the dynamics of the perturbations. 

\acknowledgments
We would like to warmly thank Christos Charmousis for discussions and for its participation at the early stage of the project. KN wants also to thank Aimeric Colléaux and David Langlois for  discussions.
NT would like to thank Shinji Mukohyama and Kazufumi Takahashi for valuable comments and discussions.

The work of EB and KN is partly supported by the French National Research Agency (ANR) via Grant No. ANR-22-CE31-0015-01 associated with the project StronG. 
KI is supported by Grant-Aid for Scientific Research from Ministry of Education, Science,
Sports and Culture of Japan (JP24K07046, JP21H05182, JP21H05189).
The work of NT was supported in part by JSPS KAKENHI Grant Numbers JP18K03623, JP21H05189, and JP22H05111. MY is supported by IBS under the project code IBS-R018-D3 and JSPS KAKENHI Grant Numbers JP21H01080.

\appendix

\section{A more detailed  analysis of the toy model}
\label{AppA}
In this appendix, we take a deeper and more detailed look at the toy model we have considered in section~\ref{sec:extended-mimetic}. It has been defined by an action of the form, 
\begin{equation}
S[g_1,g_2,\phi]= \frac12 \int dt \left(\dot {\tilde g}_1^2 + \dot {\tilde g}_2^2\right) \, ,
\label{Sg12}
\end{equation}
where $\tilde g_n$ and $g_n$ are related by  a conformal-like transformation given by
\begin{equation}
\tilde g_n = G(W)g_n~,
\qquad
W= \alpha \left(\dot g_1 g_2 - g_1 \dot g_2\right) + \beta g_1 g_2~ \,.
\label{toymodeltrsf}
\end{equation}
The functions $\alpha$ and $\beta$ depend a priori on the variables $g_n$, $\phi$ and their derivatives. However, we will limit ourselves to  specific functions which lead to interesting theories.  

\subsection{Non-invertible transformation with  $\alpha(q,\phi)$  and $\beta(q,\phi)$}
First of all, we assume the transformation \eqref{toymodeltrsf} to be non-invertible, hence 
\begin{eqnarray}
    G(W) = W^{-1/2} \, ,
\end{eqnarray}
with $W>0$. Furthermore, we assume that $\alpha$ and $\beta$ are functions of $q=g_1/g_2$ and $\phi$ only. In particular, they do not depend on $\dot \phi$ in this subsection.
Here, $q$ is the conformal-invariant variable introduced to facilitate the analysis.

In order to study the theory, it is convenient to 
reformulate the action in terms of  $q$.  For that, we first express $\tilde g_1$ and $\tilde g_2$ in terms of $q$ 
assuming $g_2>0$ for simplicity
as follows,
\begin{equation}
\tilde g_1 = q \,\tilde g_2~,
\qquad
\tilde g_2 = \frac{1}{\sqrt{ \alpha \, \dot q + \beta \, q}}~,
\label{tildeg12eq}
\end{equation}
and then we compute the action~(\ref{Sg12}), which is shown to be equivalent to the following one,
\begin{eqnarray}
\label{equivalentactiontoy}
    S[\tilde g_2,q,\phi,\pi_q] \; = \;  \int dt \, L \, ,
\end{eqnarray}
where the  Lagrangian reads
\begin{align}
L &=  
\frac{1}{2}
\left[
\left(1+q^2\right) {\dot{\tilde g}}^2_2
+ 2 q \tilde g_2 \dot q \dot{\tilde g}_2
+ {\tilde g}_2^2 \dot q^2
\right]
+ \pi_q \left(\dot q - 
\frac{1-\beta q \tilde g_2^2 }{\alpha \tilde g_2^2 }
\right)
 \\
&=
\frac{1}{2} \left[
\left(1+q^2\right) {\dot{\tilde g}}^2_2
+  \frac{2q \left(1-\beta q \tilde g_2^2 \right)}{\alpha \tilde g_2}
\dot{\tilde g}_2
+ \frac{\left(1 - \beta q \tilde g_2^2\right)^2}{\alpha^2 \tilde g_2^2}
\right]
+ \pi_q \left(\dot q - 
\frac{1-\beta q \tilde g_2^2 }{\alpha \tilde g_2^2 }
\right)\, .
\label{Stot}
\end{align}
Here, $\tilde g_2$ is viewed as an independent variable and  we have introduced a Lagrange multiplier $\pi_q$ whose equation enables us to express 
$\dot q$ in terms of  $\tilde g_2$ according to (\ref{tildeg12eq}). This is what we usually do to hide higher derivatives and to start the canonical analysis.
Notice that the Lagrange multiplier is identified as the canonical momentum of $q$ (i.e. $\pi_q = \partial L / \partial \dot q$).

Since the original action \eqref{Sg12} involves $\ddot q$, we expect  the theory to propagate an Ostrogradsky ghost. Let us see this is indeed the case with the Hamiltonian analysis of the equivalent action~\eqref{equivalentactiontoy}.
First, we compute the canonical momentum conjugate to $\tilde g_2$,
\begin{equation}
\pi = \frac{\partial L}{\partial \dot{\tilde g}_2}
=
\left(1+q^2\right){\dot{\tilde g}}_2
+  \frac{q \left(1-\beta q \tilde g_2^2 \right)}{\alpha \tilde g_2}~,
\label{pig2}
\end{equation}
and we compute the Hamiltonian 
\begin{align}
H &=
\pi \dot{\tilde g}_2
+ \pi_q \dot q
-
L
\\
&=
\frac{1}{1+q^2}\left[
\frac12 \pi^2
- \frac{q\left(1 - \beta q \tilde g_2^2\right)}{\alpha \tilde g_2} \pi
\right]
+  \pi_q \frac{\left(1 - \beta q \tilde g_2^2\right)}{\alpha \tilde g_2^2}
- \frac{\left(1 - \beta q \tilde g_2^2\right)^2}{2\left(1+q^2\right)\alpha^2 \tilde g_2^2}~.
\label{H}
\end{align}
Hence, we immediately see that the Hamiltonian is quadratic in $\pi$ but linear in $\pi_q$. This indicates that the theory admits two degrees of freedom and one of them is an Ostrogradsky ghost.

Notice that there is no kinetic term for $\phi$ in (\ref{H}), hence $\phi$ remains a non-dynamical variable unless a kinetic term for $\phi$ is introduced from the beginning in the original action~(\ref{Sg12}).

\subsection{A toy model for the extended mimetic theory}
\label{App:toymodelforextendedmimetic}
Now, we introduce a toy model corresponding to the extended mimetic theory we have introduced in the paper.
Let us recall that 
the extended mimetic theory has been defined from the non-invertible transformation given by \eqref{extmimeticgdef}
\begin{equation}
\tilde g_{\mu\nu} = 
-
X \Lambda(C_n) g_{\mu\nu}~.
\end{equation}
As $C_1$, $C_2$ and $C_3$ have been defined by \eqref{defofCalpha}, the new metric  contain terms which are schematically of the form $ g \partial^2\phi$ and $\partial g \partial \phi$,
where $g$ is a shorthand for the metric components.

In order to mimic this structure, 
we need to generalise the transformation (\ref{toymodeltrsf}) to include terms of the form $ \dot g_i \dot\phi$ and $g_i\ddot\phi$. Hence, 
we still consider  a conformal transformation  
$\tilde g_n = G(W)g_n$ where now $W$ is defined by
\begin{equation}
    W= Q\bigl(
        q,\phi,\dot\phi
    \bigr)
     \left(\dot g_1 g_2 - g_1 \dot g_2\right) 
     + P\bigl(
        q, \phi,\dot\phi,\ddot\phi
     \bigr) \,g_2^2
     = 
     g_2^2 \,\bigl(
        Q \,\dot q + P
     \bigr)
     ~,
     \label{toymodeltrsf2}
\end{equation}
with $q=g_1/g_2$. Notice that the functions $Q$ and $P$ correspond to $\alpha$ and $q\, \beta $ respectively if we compare to the notations of the previous subsection.
As $W$ still transforms according to 
$\tilde W = W G^2(W)$,
the transformation $\tilde g_n = G(W)g_n$ becomes non-invertible when $G(W) = W^{-1/2}$.

\medskip

Let us proceed with the Hamiltonian analysis of this theory. For that, we reformulate the theory and we consider the equivalent action:
\begin{eqnarray}
    S[\tilde g_2,q,\phi,A_*,B_*,\pi_q,\pi_{A_*}] \; = \; \int dt \, L \, , 
\end{eqnarray}
where now the Lagrangian is given by,
\begin{align}
    L
    &=
    \frac12 \left[
    \left(1+q^2\right) {\dot{\tilde g}}^2_2
    +  \frac{2q \left(1-\beta q \tilde g_2^2 \right)}{\alpha \tilde g_2}
    \dot{\tilde g}_2
    + \frac{\left(1 - \beta q \tilde g_2^2\right)^2}{\alpha^2 \tilde g_2^2}
    \right]
    \notag \\
    &\qquad\quad
    + \pi_q \left(\dot q - 
    \frac{1-\beta q \tilde g_2^2 }{\alpha \tilde g_2^2 }
    \right)
    + \pi_\phi \left(
        \dot\phi - A_*
    \right)
    + \pi_{A_*} \left(
        \dot A_* - B_*
    \right)
    ~,
    \label{StotE}
    \end{align}
where we introduced the variables $A_* \equiv \dot \phi$, $B_* \equiv \dot A_* = \ddot\phi$ together with the momenta $\pi_q$, $\pi_\phi$ and $\pi_{A*}$ in order to get rid of higher derivatives.
The dynamical variables for this theory are $\tilde g_2, q, \phi, A_*, B_*$.
Their canonical momenta are respectively given by $\pi$, which is the same as (\ref{pig2}), 
$\pi_q, \pi_\phi, \pi_{A_*}$ and $\pi_{B_*}$. The momentum $\pi_{B_*}$ vanishes since
\begin{equation}
    \pi_{B_*} = \frac{\partial L}{\partial \dot B_*} = 0~,
\end{equation}
and then we obtain a primary constraint.

As a consequence, the total Hamiltonian $H$ is given by
\begin{equation}
    H = H_0 + \lambda_{B_*} \pi_{B_*}~,
\end{equation}
where $\lambda_{B_*}$ enforces the primary constraint while the canonical Hamiltonian, defined by
\begin{eqnarray}
    H_0 =
    \pi \dot{\tilde g}_2 
    + \pi_q \dot q 
    + \pi_\phi \dot \phi 
    + \pi_{A_*} \dot A_* 
    + \pi_{B_*} \dot B_* 
    -L \, ,
\end{eqnarray}
is given, after a short calculation, by
\begin{align}
    H_0 &\simeq
\frac{1}{1+q^2}\left[
\frac12 \pi^2
- \frac{q\left(1 - P \tilde g_2^2\right)}{Q \tilde g_2} \pi
\right]
+ \pi_q \frac{1 - P \tilde g_2^2}{Q \tilde g_2^2} 
- \frac{1}{2\left(1+q^2\right)}
\left(\frac{1 - P \tilde g_2^2}{Q \tilde g_2}\right)^2
+ \pi_\phi A_* + \pi_{A_*} B_*~.
\end{align}

Requiring the stability of the primary constraint $\pi_{B_*}\simeq 0$ under time evolution leads to a secondary constraint $\chi_2 \simeq 0$ with
\begin{align}
    \chi_2 &\equiv \frac{d \pi_{B_*}}{d t}
    = \left\{
    \pi_{B_*}, H
    \right\}
    = - \frac{\partial H_0}{\partial B_*}
    =
    - \frac{P_{,B_*}}{1+q^2} \left(
        \frac{\tilde g_2}{Q } q \pi
        + \frac{1 - P \tilde g_2^2 }{Q^2 }
    \right)
    +\frac{P_{,B_*}}{Q} \pi_q - \pi_{A_*},
\label{constraint_chi2}
\end{align}
where $P_{,B_*}=\partial P/\partial B_*$.

This secondary constraint does not commute with the primary constraint, i.e.\ $\left\{\pi_{B_*},\chi_2\right\} \not\simeq 0$. As a consequence, requiring its stability under time evolution fixes  the Lagrange multiplier 
$\lambda_{B_*}$ and does not lead to a new constraint.

Hence, the canonical analysis stops here with  two second class constraints, and the number of  physical degrees of freedom is given by $\frac12\left(5 \times 2 - 2 \right) = 4$. Furthermore, when one eliminates
 $\pi_{A_*}$ using the constraint (\ref{constraint_chi2}), one obtains the following expression of the Hamiltonian
\begin{align}
    H_0 &\simeq
\frac{1}{1+q^2}\left(
\frac12 \pi^2
- 
\left[1 - (P -B_*P_{,B_*})\tilde g_2^2 \right]
\frac{q \pi}{Q \tilde g_2}
\right)
+
\pi_q  \frac{1 -  (P -B_*P_{,B_*}) \tilde g_2^2}{Q \tilde g_2^2} 
+   \pi_\phi A_*
\notag \\
&\quad
- \frac{\left(1 - P \tilde g_2^2\right)
\left[1 -  (P -2B_*P_{,B_*}) \tilde g_2^2\right]}{2\left(1+q^2\right)Q^2 \tilde g_2^2}~,
\end{align}
which is linear in the momenta $\pi_q$ and $\pi_\phi$. This means  that 
the theory admits two Ostrogradsky ghosts among the four degrees of freedom.

Notice that the model associated with $\alpha = \alpha(q, \phi)$ and $ \beta=\beta(q,\phi,\dot\phi)$, i.e., $Q=Q(q,\phi)$ and $ P = P(q,\phi,\dot\phi)$, can be analysed in a similar way. It turns out that such a theory admits three degrees of freedom, one of them being an Ostrogradsky ghost.

The numbers of the degrees of freedom and the presence of Ostrogradsky ghosts in the models are  summarised in the table~\ref{Table:toymodelDoF}.

\begin{table}[htbp]
  \centering
  \renewcommand{\arraystretch}{1.25}
  \begin{tabular}{|l|l|c|c|c|}
\hline
\quad~\,$\alpha$ &\qquad \,  $\beta$& \# DoF & \# ghost & corresponding theory\\
\hline\hline
$\alpha = 0$ & $\beta(q,\phi)$ & 2 DoF & 0 ghost & \\
\hline
$\alpha = 0$ & $\beta(q,\phi,\dot\phi)$ & 3 DoF & 1 ghost & standard mimetic theory\\ \hline
$\alpha(q,\phi)$ & $\beta(q,\phi)$ & 2 DoF & 1 ghost & \\ \hline
$\alpha(q,\phi)$ & $\beta(q,\phi,\dot\phi)$ & 3 DoF & 1 ghost & \\ \hline
$\alpha(q,\phi,\dot\phi)$ & $\beta(q,\phi,\dot\phi)$ & 3 DoF & 1 ghost & \\ \hline
$\alpha(q,\phi,\dot\phi)$ & $\beta(q,\phi,\dot\phi,\ddot\phi)$ & 4 DoF & 2 ghosts & extended mimetic theory\\ \hline
  \end{tabular}
  \caption{Summary of the number of the degrees of freedom (DoF) and the number of the Ostrogradsky ghosts 
  in the theories generated by applying the transformation~\eqref{toymodeltrsf} to the original theory~\eqref{Sg12}, which has two degrees of freedom without ghosts.
  In the last column, we precise the scalar-tensor theory the toy model is associated to. Note that 
  $\alpha, \beta$ correspond to $Q, P/q$ introduced in App.~\ref{App:toymodelforextendedmimetic}, and that the cases with $\alpha=0$ are studied in section~\ref{sec:toymodel_noninv-diff-trsf}.}
  \label{Table:toymodelDoF}  
\end{table}

\section{An alternative way to recover  invertible higher order conformal transformations}
\label{App:derive_C1}

In this appendix, we present an alternative way to construct invertible higher order conformal transformations that depends on
$C_1$ \eqref{explicitC1}. This is based on the method described in \cite{Babichev:2021bim}.

\medskip

To begin with, we consider a conformal transformation given by
\begin{equation}
\tilde g_{\mu\nu} = \bar \Omega(\phi, X, \mathcal{B}, Y)\, g_{\mu\nu} \,,
\label{confXBYdef}
\end{equation}
where $\mathcal{B} \equiv \nabla^\mu \nabla_\mu \phi$ and
$Y$ is defined in \eqref{defYZ}.
We also assume that the conformal factor $\bar \Omega$ is a non-vanishing function.
We apply this transformation to the FLRW ansatz,
\begin{equation}
    ds^2 
    = g_{\mu\nu} dx^\mu dx^\nu
    = -\mathfrak{n}(t) \, dt^2 + \mathfrak{a}(t)\,  d\mathbf{x}^2\,, 
    \qquad \phi = \phi(t) \,,
    \label{FLRWansatz}
\end{equation}
where we introduced $\mathfrak{n} = N^2$ and $\mathfrak{a} = a^2$.
The transformed metric takes the same FLRW form whose associated functions are now given by
\begin{equation}
 \tilde {\mathfrak{n}} = \bar \Omega \,  \mathfrak{n}\,,
 \qquad
 \tilde {\mathfrak{a}} = \bar \Omega\,  \mathfrak{a}\,.
 \label{natrsf}
\end{equation}

In the appendix G of \cite{Babichev:2021bim}, it has been shown that a transformation given by 
\begin{equation}
 \tilde {\mathfrak{n}} = F_\mathfrak{n}(\phi,X,\mathcal{B},Y) \,  \mathfrak{n}\,,
 \qquad
 \tilde {\mathfrak{a}} = F_\mathfrak{a}(\phi,X,\mathcal{B},Y)\,  \mathfrak{a}
\end{equation}
is invertible if and only if the following conditions are satisfied,
\begin{align}
&
F_{\ma,\mathcal{B}} F_{\mn,Y}
- F_{\mn,\mathcal{B}} F_{\ma,Y} = 0\,,
\label{detA=0}
\\
&
F_{\mn,\mathcal{B}}
\left[
3 F_{\mn} F_{\ma,\mathcal{B}}
- F_{\ma} 
\left(
F_{\mn,\mathcal{B}}
+ 2 X  F_{\mn,Y}
\right)
+ 3 X 
\left(
F_{\ma,X} F_{\mn,\mathcal{B}}
- F_{\ma,\mathcal{B}} F_{\mn,X}
\right)
\right]
=0
\,,
\label{nBm=0}\\
& F_\ma \neq 0
\,,
\label{firstnondeg}
\\
&
X F_{\mn,X}
+ \mathcal{B} F_{\mn,\mathcal{B}}
+ Y F_{\mn,Y} - \left(-X\right)^{3/2}
F_{\mn,\mathcal{B}}
\left(\frac{F_{\mn,Y}}{F_{\mn,\mathcal{B}}}\right)'
- F_{\mn}
\neq 0 \, ,
\label{secondnondeg}
\end{align}
together with
\begin{equation}
\left(
F_{\mn,\mathcal{B}} + 2 X F_{\mn,Y}
\right)^2
+\left(
3\mn F_{\mn,\mathcal{B}} / \ma
\right)^2
 \neq 0\,,
\quad
F_{\ma,\mathcal{B}}{}^2
+\left(
\mn F_{\mn,\mathcal{B}} / \ma
\right)^2
 \neq 0
\,.
\label{MNnonzero}
\end{equation}
In \eqref{secondnondeg}, the prime 
($'$) stands for a derivative with respect to the proper time ($f'\equiv \mn^{-1/2} \dot f$).
As mentioned below, the condition~\eqref{MNnonzero} is necessary to realise a non-trivial example of $\bar\Omega$ that depends on $\mathcal{B}$; it turns out that the conformal transformation reduces to a standard one $\bar\Omega = \bar\Omega(\phi, X)$ when this condition is not satisfied.

\medskip

The conformal transformation \eqref{natrsf} corresponds to 
the case where $F_\mn = F_\ma =\bar  \Omega(\phi,X,\mathcal{B},Y)$.
In this case, the condition \eqref{MNnonzero} reduces to $\bar \Omega_{,\mathcal{B}} \neq 0$, and
the first invertibility condition \eqref{detA=0} is automatically satisfied.
Then, the second invertibility condition \eqref{nBm=0} reduces to 
\begin{equation}
\bar \Omega_{,\mathcal{B}} - X \,\bar \Omega_{,Y}
= 0\,,
\end{equation}
which immediately implies that\footnote{In \eqref{OmegabarOmega}, we have chosen the $X$ dependence of the third argument of $\Omega$ so that it coincides with $C_1$.}
\begin{equation}
\bar \Omega(\phi,X,\mathcal{B},Y) =  \Omega\left(\phi,X, 
\frac{\mathcal{B}}{X} + \frac{Y}{X^2}
\right)
=  \Omega\left(\phi,X, 
\frac{1}{X} \left(\Box\phi + \frac{1}{ X} {\phi_\lambda X^\lambda}
\right)\right)\,.
\label{OmegabarOmega}
\end{equation}
Regarding  $\Omega$ as a function of $\phi, X$, and $C_1 = \frac1X\left(
\Box\phi + \frac{1}{ X} {\phi_\lambda X^\lambda}
\right)$,
this result coincides with the invertible higher order conformal transformation \eqref{invertibleconf} where $C_n$ is set to  $\phi$ and $C_1$.

\medskip

Plugging $
\bar \Omega(\phi,X,\mathcal{B},Y) = \Omega(\phi,X,C_1) 
$ into 
$F_\mn$ and $F_\ma$ in the last condition~\eqref{secondnondeg} for the invertibility, we find that it reduces to\footnote{To derive \eqref{secondnondegconf}, we used the fact that $\left(F_{\mn,Y} / F_{\mn,\mathcal{B}}\right)' = (1/X)' = - 
(\sqrt{\mn}/\dot \chi) Y = - Y / \sqrt{-X} $\,.
Terms containing $\Omega_{,C_1}$ appears from some terms in \eqref{secondnondeg}, but they cancel out and do not appear in the final result~\eqref{secondnondegconf}.
}
\begin{equation}
\Omega - X\,\Omega_{,X} \neq 0 \, ,
\label{secondnondegconf}
\end{equation}
which correctly reproduces the invertibility condition $\partial_X \tilde X\neq 0$ introduced around \eqref{tildeXeq} for the higher order conformal transformation~\eqref{invertibleconf}.

\medskip

To conclude, we have shown that the conformal transformation  (\ref{confXBYdef}) applied to the FRLW ansatz~\eqref{FLRWansatz} becomes invertible if and only if it is takes the form 
\begin{equation}
\tilde g_{\mu\nu} = \Omega(\phi,X,C_1) \, g_{\mu\nu}\,.
\label{invertibleconffinal}
\end{equation}
For the general background spacetime, this result gives a necessary condition for the invertibility, i.e.\ an invertible conformal transformation given by~\eqref{confXBYdef} has to take this form while there could be additional constraints to realise the invertiblity.
Of course, the invertiblity of~\eqref{invertibleconffinal} on a general background can be easily proven as shown in section~\ref{sec:higherorderconf}. Hence, \eqref{invertibleconffinal} is the most general invertible transformation within the class of transformations given by \eqref{confXBYdef}.

\medskip

We could generalise the above analysis to cases where the conformal factor $\Omega$ depends also on other scalar quantities, such as $Z$ \eqref{defYZ}. However, a short calculation shows that $Z = Y^2/X$ in a FLRW spacetime~\eqref{FLRWansatz}, and then the dependence on $Z$ degenerates with that on $X$ and $Y$.
Similar degeneracy occurs for other higher-order scalars in this ansatz, like 
\begin{eqnarray}
    {\phi}_{\mu}^{\nu} \, {{\phi}}^{\mu}_{\nu} = \left(\frac{Y}{2X}\right)^2 + \frac13 \left(\frac{Y}{2X} - \mathcal{B}\right)^2 \, .
\end{eqnarray} 
Hence, to generalise this approach by taking into account the dependency on higher order scalars, 
one would need to consider less symmetric background spacetimes.

\section{Hamiltonian analysis of  mimetic theories in a FLRW spacetime}
\label{AppB}
In this section, we give some details of the Hamiltonian analysis of the  extended mimetic theories in the context of cosmology. The spacetime is described by a FLRW metric and the scalar field depends on time only,
\begin{equation}
\tilde g_{\mu\nu} dx^\mu dx^\nu = -\tilde N^2 dt^2 + \tilde a^2 d\mathbf{x}^2\, ,
\qquad
\phi = \phi(t)~\,.
\end{equation}
For completeness, we also reproduce the analysis of the standard mimetic theory. 

\subsection{Extended mimetic theory}
We consider a non-invertible conformal transformation given by
\begin{equation}
\tilde g_{\mu\nu} = -X \Lambda(C_1) g_{\mu\nu}~,
\end{equation}
where $C_1$ is the conformal-invariant scalar, i.e.\ $C_1 = \tilde C_1$, given by,
\begin{equation}
C_1[g,\phi] = \hat \square \phi
= \frac1X \left(
\square \phi + \frac1X \phi_\lambda X^\lambda
\right)~ \, .
\end{equation}
In that case, we have,
\begin{equation}
\tilde X \equiv \tilde g^{\mu\nu} \phi_\mu \phi_\nu = - \frac{1}{\Lambda(C_1)}~,
\end{equation}
which illustrates the non-invertibility of the transformation since one cannot express $X$ in terms of~$\tilde X$.

\medskip

The extended mimetic action is defined from this transformation as follows,
\begin{equation}
S[g_{\mu\nu}, \phi] = S_{EH}[\tilde g_{\mu\nu}]~,
\label{Sdef}
\end{equation}
where $S_{EH}$ is the Einstein-Hilbert action.
 This action involves higher derivatives of the field. As usual, in order to perform its analysis, we consider instead the equivalent action,
\begin{equation}
S = \int d^4x \sqrt{-\tilde g}\left\{
\tilde R 
+ \lambda_X\left( \tilde X
+
\frac1{\Lambda}\right) 
+ \lambda_{C_1}\left[
 C_1 -  \frac1{\tilde X} \left(
\tilde \square \phi + \frac1{\tilde X} \phi_\lambda \tilde X^\lambda
\right)
\right]
\right\}~,
\end{equation}
where $\lambda_X$ and $\lambda_{C_1}$ fix $\tilde X$ and $C_1$ in terms of the metric components  and the scalar field $\phi$. 

\medskip

When the metric is restricted to a FLRW form (\ref{FRW}), the Lagrangian reduces to 
\begin{align}
-\frac{6\tilde a \dot {\tilde a}^2}{\tilde N}
+  \frac{\lambda_X \tilde a^3}{\tilde N\Lambda }
\left(
    \tilde N^2 -\Lambda  A_*^2 
\right)
+  \tilde N\tilde a^3 \lambda_{C_1}\left[
 C_1 - 
\frac{3}{ A_*^2}
\left(
\dot { A}_*
+ 
\frac{ A_*\dot {\tilde a} }{\tilde a} - \frac{ A_*\dot {\tilde N}}{\tilde N}
\right)
\right]
+ 
\pi_\phi \left(\dot\phi -  A_* \right) \, ,
\end{align}
up to irrelevant boundary terms. 
Notice that we have also introduced the Lagrange multiplier 
$\pi_{\phi}$
in order to replace formally $\dot\phi$ by $ A_*
$ and then to eliminate higher derivatives of $\phi$ from the action.
To lighten the notations, we omit to mention the tilde of the metric coefficients (e.g.\ $\tilde a \to a$).

\medskip

The theory involves seven dynamical variables ($N,a, \phi, A_*, C_1, \lambda_X, \lambda_{C_1}$). The conjugate momenta are given by,
\begin{equation}
\pi_a = 
-a\left(
\frac{12\dot a}{N}
+\frac{3Na\lambda_{C_1}}{A_*}
\right)
\,,
\quad
\pi_N = \frac{3a^3\lambda_{C_1}}{A_*}\,,
\quad
\pi_{A_*} = -\frac{3Na^3 \lambda_{C_1}}{A_*^2}\,,
\quad
\pi_{C_1} = \pi_{\lambda_X} = \pi_{\lambda_{C_1}} = 0~,
\label{Pconst_FRW}
\end{equation}
and the momentum conjugate to $\phi$ is $\pi_\phi$.
The first relation enables us to express  $\dot a$ in terms of $\pi_a$, while the others are  primary constraints. 
Therefore, the total Hamiltonian is given by
\begin{equation}
H = H_0 
+ \kappa_N \left(
\pi_N - \frac{3a^3\lambda_{C_1}}{A_*}
\right)
+ \kappa_{A_*} \left(
\pi_{A_*} + \frac{3Na^3 \lambda_{C_1}}{A_*^2}
\right)
+ \kappa_{C_1} \pi_{C_1}
+ \kappa_{\lambda_X} \pi_{\lambda_X}
+ \kappa_{\lambda_{C_1}} \pi_{\lambda_{C_1}}\, ,
\end{equation}
where the canonical Hamiltonian $H$ reads
\begin{align}
H_0 &\simeq
-\frac{N}{24a}\pi_a^2 
- \frac{N^2 a \lambda_{C_1}}{4 A_*} \pi_a
+ \pi_\phi A_* 
+ a^3 \left(
- N \lambda_{C_1} C_1
- \frac{3N^3 \lambda_{C_1}^2}{8 A_*^2}
+ \frac{\lambda_X A_*^2}{N}
- \frac{N \lambda_X}{\Lambda}
\right)
\, .
\label{H_ex-mimetic_FRW}
\end{align}

The stability under time evolution of the primary constraints $\pi_{C_1} \simeq 0$ and $\pi_{\lambda_{X}}\simeq  0$ leads to the following secondary constraints,
\begin{align}
&\frac{d\pi_{C_1}}{dt} = \left\{\pi_{C_1}, H\right\}
= N a^3 \left(
\lambda_{C_1} - \lambda_X\frac{\Lambda'}{\Lambda^2} 
\right)
\simeq 0
\,, \\
&\frac{d\pi_{\lambda_X}}{dt} = \left\{\pi_{\lambda_X}, H\right\}
=
a^3\left(
\frac{N}{\Lambda} - \frac{A_*^2}{N}
\right)
\simeq 0\,,
\label{Sconst_FRW(N)}
\end{align}
while the stability  of 
$\pi_{A_*} + \frac{3Na^3 \lambda_{C_1}}{A_*^2}\simeq 0$
and $ \pi_{\lambda_{C_1}}\simeq 0$ fixes the Lagrange multipliers 
$\kappa_{\lambda_{C_1}}$
and
$\kappa_{{A_*}}$.
It is interesting to notice that the secondary constraint (\ref{Sconst_FRW(N)}) corresponds to the constraint of the extended mimetic theory $\Lambda \tilde X = -1 $.

Finally, the stability of $\pi_N - \frac{3a^3\lambda_{C_1}}{A_*} \simeq 0$ leads to one more secondary constraint, which could be identified to the Hamiltonian constraint,
\begin{equation}
\frac{\pi_a^2}{24 a} + \frac{N a \lambda_{C_1}}{4 A_*} \pi_a
- \frac{1}{N} A_* \pi_\phi
+ a^3\left(
\lambda_{C_1} C_1 + \frac{3 N^2 \lambda_{C_1}^2}{8 A_*^2}
- \frac{\lambda_X A_*^2}{N^2} + \frac{\lambda_X}{\Lambda}
\right)
\simeq 0 \,. \\
\label{Hconst}
\end{equation}
One shows that the canonical analysis stops here with no more constraints. All the remaining stability conditions lead to  fixing the Lagrange multipliers
except for $\kappa_N$, which remains arbitrary.

We end up with 8 constraints from which we can extract (with suitable linear combinations) 2~first class constraints. One of these has the simple expression
\begin{equation}
\pi_{A_*} + \frac{N}{A_*} \pi_N \simeq 0\,,
\label{firstclass1}
\end{equation}
while the expression of the other one is rather cumbersome and takes the form,
\begin{multline}
\frac{\pi_a^2}{24 a} + \frac{N a \lambda_{C_1}}{4 A_*} \pi_a
- \frac{1}{N} A_* \pi_\phi
+ a^3\left(
\lambda_{C_1} C_1 + \frac{3 N^2 \lambda_{C_1}^2}{8 A_*^2}
- \frac{\lambda_X A_*^2}{N^2} + \frac{\lambda_X}{\Lambda}
\right)
\\
+ c_{N} \left(
\pi_N - \frac{3a^3\lambda_{C_1}}{A_*}
\right)
+ c_{C_1} \pi_{C_1}
+ c_{{\lambda_X}} \pi_{\lambda_X}
+ c_{{\lambda_{C_1}}} \pi_{\lambda_{C_1}}
\simeq 0\,,
\label{firstclass2}
\end{multline}
where $c_{N},c_{{C_1}},c_{{\lambda_X}},c_{{\lambda_{C_1}}}$ are  functions whose expression is not needed.

We show that the remaining 6 constraints are second class.
Then, the number of the physical degrees of freedom is given by $\frac12 (7 \times 2 - 2\times 2  - 6 ) = 2$.

\medskip

One can simplify the expression of the Hamiltonian 
(\ref{H_ex-mimetic_FRW})
which, after using some of the second class constraints,  takes the form, 
\begin{equation}
H_0
\simeq
-\frac{N}{24a^3} (a \pi_a + N\pi_N)^2 + \pi_\phi A_* + \frac{A_* }{3N \Lambda'(C_1)} 
\left(
N^2 \Lambda  - \Lambda^2 A_*^2 + N^2 C_1 \Lambda' 
\right)
\pi_N
\, .
\label{H_extendedmimetic}
\end{equation}
Furthermore, the constraint (\ref{Sconst_FRW(N)}) can be used to replace $N$ by $N \simeq \pm A_*/\sqrt{\Lambda}$. Using this relation with a plus sign without loss of generality,
(\ref{H_extendedmimetic}) can be simplified to
\begin{equation}
H_0 \simeq 
N\left[
-\frac{1}{24a^3}
\left(
a \pi_a + \sqrt{\Lambda} A_* \pi_N
\right)^2
- \frac{A_* C_1}{3} \pi_N
+ \frac{1}{\sqrt{\Lambda}} \pi_\phi
\right]~.
\end{equation}
This is linear in $\pi_N$ and $\pi_\phi$ while quadratic in $\pi_a$ (shifted by a $\pi_N$ term).
The expression in the square brackets coincides with the Hamiltonian constraint (\ref{Hconst}) up to the overall sign on shell.
Also, using the constraint (\ref{firstclass1}), we obtain,
\begin{equation}
\label{HFRLWextended}
H_0 \simeq 
N\left[
-\frac1{24a^3} \left(
a \pi_a - A_* \pi_{A_*}
\right)^2
+\frac{1}{3\sqrt{\Lambda}} \left(
A_* C_1 \pi_{A_*} + 3 \pi_\phi
\right)
\right]
~.
\end{equation}
It is interesting to compare the Hamiltonian of the extended mimetic theory with the Hamiltonian~(\ref{H_standardmimetic}) of the standard mimetic theory, which is shown to be  linear in one momentum only. 
The fact that \eqref{HFRLWextended} is linear in two momenta  may suggest that the  extended mimetic theory is plagued with one more Ostrogradsky ghost than the standard mimetic theory, even though this needs a careful analysis to be confirmed.

\subsection{Standard mimetic theory}
For  completeness, let us recall the analysis of the standard mimetic theory restricted to a cosmological metric. 
The action of the full theory is given by 
\begin{equation}
S = \int d^4x \sqrt{-\tilde g}\left[
\tilde R 
+ \lambda_X\bigl( \tilde X + 1\bigr)
\right]~,
\end{equation}
which is equivalent to the theory generated from the Einstein-Hilbert action for the metric $\tilde g_{\mu\nu}$ with a non-invertible conformal transformation $\tilde g_{\mu\nu} = -X g_{\mu\nu}$ applied.
In the case of  a FLRW spacetime, its Lagrangian can be reformulated as follows,
\begin{equation}
L= -6 \frac{a \dot a^2}{N} +\lambda_X \frac{a^3  }{N} \left(
N^2 - A_*^2
\right)
+ \pi_\phi \bigl( \dot \phi - A_*  \bigr)
\, ,
\end{equation}
up to irrelevant boundary terms, 
where we have introduced a new variable $A_* \equiv \dot \phi$, as usual. 

The variable $\pi_\phi$ is clearly conjugate to $\phi$, and
the  momentum conjugate to $a$ is easily computed and is given by 
$\pi_a = -12 a \dot a / N$. All the remaining momenta vanishes,
\begin{eqnarray}
    \pi_N \simeq 0 \,, \qquad 
    \pi_{\lambda_X}\simeq 0 \, , \qquad  
    \pi_{A_*}\simeq 0 \, ,
\end{eqnarray}
and therefore the theory admits three primary constraints.

As a consequence, the total Hamiltonian is given by
\begin{equation}
H =
H_0  
+ \kappa_N \pi_N
+ \kappa_{\lambda_X} \pi_{\lambda_X}
+ \kappa_{A_*} \pi_{A_*}
\, ,
\end{equation}
 where the canonical Hamiltonian reads,
\begin{equation}
H_0
=\pi_a \dot a + \pi_\phi \dot \phi - L
=
- \frac{N}{24 a}\pi_a^2
+ \lambda_X \frac{a^3}{N} \left(A_*^2 - N^2\right) 
+ \pi_\phi A_*
\,.
\end{equation}

Requiring the stability under time evolution of these primary constraints leads to three new constraints whose expressions are,
\begin{eqnarray}
\frac{d \pi_N}{d t} 
&=&
\frac{a^3\lambda_X}{N^2}
\left(
N^2 + A_*^2
\right)
+ \frac{\pi_a^2}{24 a}
\simeq 0\,, \nonumber \\
\frac{d \pi_{\lambda_X}}{d t} 
&=&
\frac{a^3}{N}\left(
N^2 - A_*^2
\right)
\simeq 0\, , \label{Sconst_FRW_mimetic(N)} \\
\frac{d \pi_{A_*}}{d t}
&=&
-\pi_\phi - \frac{2 a^3A_* }{N}  \lambda_X
\simeq 0\, . \nonumber
\end{eqnarray}
Notice that the first of these secondary constraints is in fact  the Hamiltonian constraint. The stability of these constraints fixes the Lagrange multipliers $\kappa_{\lambda_X}$ and $\kappa_{A*}$ while $\kappa_N$ remains free. 
Finally, the Hamiltonian analysis stops here with 6 constraints.

One can show that the following two combinations of constraints are first class,
\begin{equation}
\pi_{A_*} + \pi_N \simeq  0\,,
\qquad
\frac{a^3\lambda_X}{N^2}
\left(
N^2 + A_*^2
\right)
+ \frac{\pi_a^2}{24 a}
-\pi_\phi - \frac{2 a^3 A_* \lambda_X}{N}
+ \frac{\sqrt{\frac32}\pi_{\lambda_X} \pi_\phi^{3/2}}{2a^{9/2}}
\simeq 0\,.
\label{firstclass_standardmimetic}
\end{equation}
The other constraints are all second class, and hence the number of the physical degrees of freedom is given by $\frac12 (5\times 2 - 2 \times 2 - 4) = 1$. 

Furthermore, when one eliminates $\lambda_X$ and $A_*$ using (\ref{Sconst_FRW_mimetic(N)}), the Hamiltonian becomes
\begin{equation}
H_0 \simeq N \left(-\frac{\pi_a^2}{24a }  + \pi_\phi  \right)
\, .
\label{H_standardmimetic}
\end{equation}
It vanishes on shell and is  linear in the momentum $\pi_\phi$. The question whether this implies that the theory admits an Ostrogradsky ghost is subtle and has been discussed in the literature ~\cite{Barvinsky:2013mea,Golovnev:2013jxa,Chaichian_2014,Langlois:2018jdg}.

\bibliographystyle{utphys}
\bibliography{Biblio_Mimetic}

\end{document}